\documentclass[12pt]{article}
\usepackage{amsmath,amssymb,pictex,cite,eepic,epsfig,psfrag,url}
\advance \topmargin by -\headheight
\advance \topmargin by -\headsep     
\evensidemargin \oddsidemargin
\makeatletter
\@addtoreset{equation}{section}

\makeatother
\def\Maketitle{{\def\newpage{}\maketitle}}
\makeatletter
\def\Appendix{\appendix
  \def\@seccntformat##1{Appendix~\csname the##1\endcsname.~~}}
\makeatother
\def\sn{\mathop{\rm sn}\nolimits}

\def\Xint#1{\mathchoice
{\XXint\displaystyle\textstyle{#1}}%
{\XXint\textstyle\scriptstyle{#1}}%
{\XXint\scriptstyle\scriptscriptstyle{#1}}%
{\XXint\scriptscriptstyle\scriptscriptstyle{#1}}%
\!\int}
\def\XXint#1#2#3{{\setbox0=\hbox{$#1{#2#3}{\int}$}
\vcenter{\hbox{$#2#3$}}\kern-.5\wd0}}

\def\dashint{\Xint-}
\begin{document}
\title{\textbf{Differential equation for four-point\\ correlation function in Liouville field theory\\                    and elliptic four-point conformal blocks}}
\author{Vladimir A.~Fateev$^{1,2}$, Alexey V.~Litvinov$^{1,3}$, Andr\'e Neveu$^{2}$ and Enrico Onofri$^{4}$
\\[\bigskipamount]$^1$
~\parbox[t]{0.85\textwidth}{\normalsize\it\raggedright
Landau Institute for Theoretical Physics,
142432 Chernogolovka, Russia}\\[\medskipamount]
$^2$~\parbox[t]{0.85\textwidth}{\normalsize\it\raggedright
Laboratoire de Physique Th\'eorique et Astroparticules, UMR5207 CNRS-UM2, Universit\'e
Montpellier~II, Pl.~E.~Bataillon, 34095 Montpellier, France}\\[1.6\bigskipamount]
$^3$~\parbox[t]{0.85\textwidth}{\normalsize\it\raggedright
NHETC, Department of Physics and Astronomy,
Rutgers University,\\ Piscataway, NJ 08854-0819, USA}
\\[1.6\bigskipamount]
$^4$~\parbox[t]{0.85\textwidth}{\normalsize\it\raggedright
Dipartimento di Fisica, Universit\`a di Parma, and I.N.F.N. Gruppo di Parma, 43100 Parma, Italy}
}
\date{}
\rightline{RUNHETC-2008-27}
\Maketitle
\begin{abstract}
Liouville field theory on a sphere is considered. We explicitly derive a differential equation for four-point correlation functions with one degenerate field $V_{-\frac{mb}{2}}$. We introduce and study also a class of four-point conformal blocks which can be calculated exactly and represented by finite dimensional integrals of elliptic theta-functions for arbitrary intermediate dimension. We study also the bootstrap equations for these conformal blocks and derive integral representations for corresponding four-point correlation functions. A relation between the one-point correlation function of a primary field on a torus and a special four-point correlation function on a sphere is proposed.
\end{abstract}
\section{Introduction}
Liouville field theory has attracted a lot of attention since the discovery of its application in the quantization of the string in non-critical dimension in 1981 \cite{Polyakov:1981rd}. It gives an important example of a non-rational unitary conformal field theory (CFT) with central charge $c>25$.  From another side any correlation function in minimal models i.~e. in rational CFT's with central charge $c<1$ can be obtained from the appropriate correlation functions in LFT by analytic continuation\footnote{At this point some care is needed (see Ref \cite{Zamolodchikov:2005fy}).}. 

Liouville field theory is defined on a two-dimensional surface with metric $\hat{g}_{ab}$ by the local Lagrangian density
\begin{equation}\label{Lagrangian}
   \mathcal{L}=\frac{1}{4\pi}\hat{g}^{ab}\partial_a\varphi\partial_b\varphi+\mu e^{2b\varphi}+
   \frac{Q}{4\pi}\hat{R}\,\varphi,
\end{equation}
where $\hat{R}$ is associated curvature. This theory is conformally invariant if the coupling constant $b$ is related with the background charge $Q$ as
\begin{equation}
    Q=b+\frac{1}{b}.
\end{equation}
The symmetry algebra of this conformal field theory is the Virasoro algebra 
\begin{equation}\label{Virasoro}
    [L_m,L_n]=(m-n)L_{m+n}+\frac{c_L}{12}(n^3-n)\delta_{n,-m}
\end{equation}
with central charge
\begin{equation}
    c_L=1+6Q^2.
\end{equation}
Primary fields $V_{\alpha}$ in this theory, which are associated with exponential fields $e^{2\alpha\varphi}$, have conformal dimensions
\begin{equation}
   \Delta(\alpha)=\alpha(Q-\alpha).
\end{equation}
The fields $V_{\alpha}$ and $V_{Q-\alpha}$ have the same conformal dimensions and represent the same primary field, i.e. they are proportional to each other:
\begin{equation}\label{Reflection}
   V_{\alpha}=R(\alpha)V_{Q-\alpha},
\end{equation}
with the function
\begin{equation*}
  R(\alpha)=\frac{(\pi\mu\gamma(b^2))^{(Q-2\alpha)/b}}{b^2}\frac{\gamma(2b\alpha-b^2)}
  {\gamma(2-2\alpha/b+1/b^2)},
\end{equation*}
known as the reflection amplitude. Here and later we use the notation
\begin{equation}
\gamma(x)=\Gamma(x)/\Gamma(1-x).
\end{equation}
Since Liouville field theory is conformal invariant it can be solved exactly at least in principle. 
In particular it means that one can find explicitly correlation functions of all local fields. Following the ideas of the operator product expansion (OPE) it is sufficient to find the structure constants of the OPE in order to find all correlation functions in the theory. In the case of Liouville field theory this problem simplifies drastically due to conformal invariance of the theory \cite{Belavin:1984vu}. Namely, the operator product expansion of two primary fields $V_{\alpha_1}$ and  $V_{\alpha_2}$ has the form
\begin{equation}\label{OPE}
  V_{\alpha_1}(z)V_{\alpha_2}(0)=\frac{1}{2}\int_{-\infty}^{\infty}
  C_{\alpha_1\alpha_2}^{\frac{Q}{2}+iP}z^{\frac{Q^2}{4}+P^2-\Delta_1-\Delta_2}
  \Bigl[V_{\frac{Q}{2}+iP}(0)+\dots\Bigr]dP.
\end{equation}
Here by $\dots$ we denote the contribution of the so-called descendant fields i.e. fields which can be obtained by the action of the negative part of the Virasoro algebra \eqref{Virasoro} on the primary fields; this contribution is universal and completely determined by conformal invariance \cite{Belavin:1984vu}. The part of the operator product expansion which is not fixed by conformal invariance is the set of structure constants of primary fields $C_{\alpha_1\alpha_2}^{\alpha_3}$. They can be found from the condition of associativity of the operator algebra  \cite{Teschner:1995yf}. It is better to write down the three-point correlation function
\begin{equation}\label{3point-def}
  \langle V_{\alpha_1}(z_1)V_{\alpha_2}(z_2)V_{\alpha_3}(z_3)\rangle=\frac{C(\alpha_1,\alpha_2,\alpha_3)}
  {|z_{12}|^{2\Delta_1+2\Delta_2-2\Delta_3}|z_{13}|^{2\Delta_1+2\Delta_3-2\Delta_2}
  |z_{23}|^{2\Delta_2+2\Delta_3-2\Delta_1}},
\end{equation}
which is trivially related to the structure constant $C_{\alpha_1\alpha_2}^{\alpha_3}$ as
\begin{equation}
    C_{\alpha_1\alpha_2}^{\alpha_3}=C(\alpha_1,\alpha_2,Q-\alpha_3).
\end{equation}
The constant $C(\alpha_1,\alpha_2,\alpha_3)$ introduced by \eqref{3point-def} was proposed in \cite{Dorn:1992at,Dorn:1994xn} and passed a lot of crucial tests in \cite{Zamolodchikov:1995aa} (more rigorous proof can be found in  Ref. \cite{Teschner:2003en})
\begin{equation}\label{3point}
    C(\alpha_1,\alpha_2,\alpha_3)=
    \Bigl[\pi\mu\gamma(b^2)b^{2-2b^2}\Bigr]^{\frac{(Q-\alpha)}{b}}
    \frac{\Upsilon(b)\Upsilon(2\alpha_1)\Upsilon(2\alpha_2)\Upsilon(2\alpha_3)}
    {\Upsilon(\alpha-Q)\Upsilon(\alpha-2\alpha_1)\Upsilon(\alpha-2\alpha_2)
     \Upsilon(\alpha-2\alpha_3)},
\end{equation}
where $\alpha=\alpha_1+\alpha_2+\alpha_3$ and $\Upsilon(x)$ is an entire function, selfdual with respect to $b\rightarrow b^{-1}$ satisfying functional relations 
\begin{equation}
 \begin{aligned}
   &\Upsilon(x+b)=\gamma(bx)b^{1-2bx}\Upsilon(x),\\
   &\Upsilon(x+b^{-1})=\gamma(b^{-1}x)b^{2b^{-1}x-1}\Upsilon(x),
 \end{aligned}
\end{equation}
which for general real values of the parameter $b^2$ have a unique solution given by the integral \eqref{Upsilon-Integral} with normalization $\Upsilon(Q/2)=1$. The analytic part of the  operator product expansion \eqref{OPE} i.~e. the contribution of descendant fields contains more non-triviality. Despite the fact that this contribution is determined completely by conformal invariance its determination represents a tedious problem. This problem was attacked in Ref.s \cite{Zamolodchikov:1985ie,Apikyan:1987mn,Zamolodchikov:1985tmf} where analytical properties of the so-called four-point conformal block were studied. The four-point conformal block is defined as a contribution of the particular primary field together with its descendants in four-point correlation function. Namely, due to \eqref{OPE} the four-point correlation function can be represented as a sum over intermediate states
\begin{multline}\label{OPE-4point}
 \langle V_{\alpha_1}(z_1,\bar{z}_1)V_{\alpha_2}(z_2,\bar{z}_2)
         V_{\alpha_3}(z_3,\bar{z}_3)V_{\alpha_4}(z_4,\bar{z}_4)\rangle=
  \prod_{i<j}|z_i-z_j|^{2\gamma_{ij}}\times\\\times
  \frac{1}{2}\int\limits_{\mathcal{C}}C\Bigl(\alpha_1,\alpha_2,\frac{Q}{2}+iP\Bigr)
  C\Bigl(\frac{Q}{2}-iP,\alpha_3,\alpha_4\Bigr)
  \biggl|\mathfrak{F}_P
   \Bigl(\genfrac{}{}{0pt}{}{\alpha_2\;\;\alpha_3}{\alpha_1\;\;\alpha_4}\Bigl|x\Bigr)\biggr|^2
   dP,
\end{multline}
here $\gamma_{ij}$ are standard combinations of conformal dimensions \cite{Belavin:1984vu} and $x$ is the anharmonic ratio for the four points $z_j$:
\begin{equation}
  x=\frac{z_{12}z_{34}}{z_{13}z_{24}}.
\end{equation}
The contour of integration $\mathcal{C}$ in \eqref{OPE-4point} goes along the real axis with possible deformation due to the presence of discrete terms (see Ref. \cite{Zamolodchikov:1995aa} for details). The conformal block $\mathfrak{F}_P\Bigl(\genfrac{}{}{0pt}{}{\alpha_2\;\;\alpha_3}{\alpha_1\;\;\alpha_4}\Bigl|x\Bigr)$ sums up all the intermediate descendant states of a given primary one with conformal dimension $\Delta=\frac{Q^2}{4}+P^2$ in the operator product expansion \eqref{OPE-4point}. T-channel conformal block usually is represented graphically as
\begin{equation}\label{Conf.Block-grafical-definition}
\mspace{-240mu}
\mathfrak{F}_P\Bigl(\genfrac{}{}{0pt}{}{\alpha_2\;\;\alpha_3}{\alpha_1\;\;\alpha_4}\Bigl|x\Bigr)=
  \begin{picture}(0,60)(-100,38)
    \Thicklines
    \unitlength 2pt 
    \put(20,20){\line(4,3){20}}
    \put(35,37){\mbox{$(\alpha_3, \infty)$}}
    \put(20,20){\line(4,-3){20}}
    \put(35,0){\mbox{$(\alpha_4,1)$}}
    \put(-50,37){\mbox{$(\alpha_2, 0)$}}
    \put(20,20){\line(-4,0){40}}
    \put(-50,0){\mbox{$(\alpha_1,x)$}}
    \put(-20,20){\line(-4,3){20}}
    \put(-20,20){\line(-4,-3){20}}
    \put(-10,25){\mbox{$P^2+\frac{Q^2}{4}$}}
  \end{picture}
  \vspace{17mm}  
\end{equation}
Unfortunately this function is not known in a closed form, but it can be found as a power series expansion. The most efficient way to do that was proposed in Ref. \cite{Zamolodchikov:1985tmf}. According to \cite{Zamolodchikov:1985tmf} we introduce instead of usual coordinate $x$ a new coordinate $\tau$
\begin{equation}\label{tau}
     \tau=i\,\frac{K(1-x)}{K(x)},
\end{equation}
which is the coordinate on the upper half plane ($\text{Im}(\tau)>0$). Here $K(x)$ is the elliptic integral of the first kind
\begin{equation}
  K(x)=\frac{1}{2}\int_0^1\frac{dt}{\sqrt{t(1-t)(1-xt)}}.
\end{equation}
The elliptic conformal block $\mathfrak{H}_P\Bigl(\genfrac{}{}{0pt}{}{\alpha_2\;\;\alpha_3}{\alpha_1\;\;\alpha_4}\Bigl|q\Bigr)$ is defined as follows
\begin{equation}\label{Elliptic-Block-definition}
   \mathfrak{F}_P\Bigl(\genfrac{}{}{0pt}{}{\alpha_2\;\;\alpha_3}{\alpha_1\;\;\alpha_4}\Bigl|x\Bigr)=
   (16q)^{P^2}x^{\frac{Q^2}{4}-\Delta_1-\Delta_2}(x-1)^{\frac{Q^2}{4}-\Delta_1-\Delta_4}
    \theta_3(q)^{3Q^2-4\sum_k\Delta_k}
   \mathfrak{H}_P\Bigl(\genfrac{}{}{0pt}{}{\alpha_2\;\;\alpha_3}{\alpha_1\;\;\alpha_4}\Bigl|q\Bigr),
\end{equation}
where $q=e^{i\pi\tau}$ and $\theta_3(q)$ is the theta constant (see definition in appendix \ref{formulae}). The function $\mathfrak{H}_P\Bigl(\genfrac{}{}{0pt}{}{\alpha_2\;\;\alpha_3}{\alpha_1\;\;\alpha_4}\Bigl|q\Bigr)$ satisfies a nice recursive relation (see Ref. \cite{Zamolodchikov:1985tmf} and also Ref. \cite{Zamolodchikov:2006cx}) which leads, in particular, to an effective algorithm for calculation of its expansion in power series of $q$
\begin{equation}\label{H-expansion}
   \mathfrak{H}_P\Bigl(\genfrac{}{}{0pt}{}{\alpha_2\;\;\alpha_3}{\alpha_1\;\;\alpha_4}\Bigl|q\Bigr)
   =1+\sum_{L=1}^{\infty}
   \mathfrak{H}_P^{(L)}\Bigl(\genfrac{}{}{0pt}{}{\alpha_2\;\;\alpha_3}{\alpha_1\;\;\alpha_4}\Bigr)q^L.
\end{equation}
Due to the  much better convergence of this sum than in the usual variable $x$ the representation of the conformal block in terms of the elliptic variable $q$ is more convenient for numerical studies. The coefficients $\mathfrak{H}_P^{(L)}\Bigl(\genfrac{}{}{0pt}{}{\alpha_2\;\;\alpha_3}{\alpha_1\;\;\alpha_4}\Bigr)$ are rational functions of $P^2$
\begin{equation}
 \begin{gathered}
    \mathfrak{H}_P^{(1)}\Bigl(\genfrac{}{}{0pt}{}{\alpha_2\;\;\alpha_3}{\alpha_1\;\;\alpha_4}\Bigr)=
    \frac{32b^2(\left(\alpha_1-\frac{Q}{2}\right)^2-\left(\alpha_2-\frac{Q}{2}\right)^2)
    (\left(\alpha_3-\frac{Q}{2}\right)^2-\left(\alpha_4-\frac{Q}{2}\right)^2)}{4P^2b^2+(b^2+1)^2},\\
    \dots\dots\dots\dots\dots\dots\dots\dots\dots 
 \end{gathered}
\end{equation}
In principle, using the recursive relations any coefficient $\mathfrak{H}_P^{(L)}\Bigl(\genfrac{}{}{0pt}{}{\alpha_2\;\;\alpha_3}{\alpha_1\;\;\alpha_4}\Bigr)$ in expansion \eqref{H-expansion} can be found explicitly, but they are rather cumbersome for large values of $L$ (even for $L=2$ the answer is very complicated). In order to find the four-point correlation function \eqref{OPE-4point} one has to perform the integration over the intermediate momentum $P$ in \eqref{OPE-4point}. Altogether it represents a complicated numerical problem, which was studied for particular cases in \cite{Zamolodchikov:1995aa,Zamolodchikov:2006cx,Zamolodchikov:2005jb}. In some cases the four-point correlation function can be found in explicit form. In particular, if one of the fields is degenerate (see below) the correlation function satisfies a differential equation of Fuchsian type with solutions having explicit integral representations.

This paper consists of two parts. The first part (section \ref{DiffEq}) is devoted to the derivation of the explicit form of the differential equation for the four-point correlation functions in Liouville field theory. In the second part of this paper (section \ref{Int-potentials}) we discover a family of conformal blocks which have explicit integral representations for arbitrary value of intermediate momentum $P$. We study the bootstrap conditions for these conformal blocks and derive the four-point correlation functions. We propose also a relation between one-point correlation function on a torus and special four-point correlation function on a sphere in LFT. In appendixes we collect some formulae used in this paper and give proofs of some statements.

The content of this paper is closely related with the subject where the seminal contribution of Alyosha Zamolodchikov has a great importance. The authors had  a chance to know him personally and could estimate his scientific virtuosity, originality and creativity. We dedicate this paper to the memory of Alyosha.  
\section{Differential equation for four-point correlation function}\label{DiffEq}
Among the primary fields $V_{\alpha}$ the so-called degenerate fields are of special interest. They form a closed subalgebra of the operator algebra and are characterized by the parameters
\begin{equation}\label{alpha-mn}
  \alpha=\alpha_{mn}=-\frac{mb}{2}-\frac{n}{2b}.
\end{equation}
Any degenerate field $V_{\alpha_{mn}}$ has a null-vector  in its Verma module at level $(m+1)(n+1)$.
In particular the primary field $V_{-\frac{b}{2}}$ has a null-vector at the second level
\begin{equation}\label{F12-null_vector_condition}
    \left(L_{-1}^2+b^2L_{-2}\right)V_{-\frac{b}{2}}=0.
\end{equation}
As a consequence of the null-vector condition \eqref{F12-null_vector_condition} the correlation function which contains this degenerate field and an arbitrary number of general primary fields\footnote{Here and later we will drop sometimes the dependence on the antiholomorphic variables $\bar{z}$.}
\begin{equation}\label{n-point-with-Phi12}
  \langle V_{-\frac{b}{2}}(z)V_{\alpha_1}(z_1)\dots V_{\alpha_n}(z_n)\rangle
\end{equation}
satisfies the second order partial differential equation \cite{Belavin:1984vu}
\begin{equation}\label{Phi12-general-equation}
  \left[\partial_z^2+b^2\left(\sum_{k=1}^n\frac{\Delta(\alpha_k)}{(z-z_k)^2}+
  \frac{\partial_k}{(z-z_k)}\right)\right]
  \langle V_{-\frac{b}{2}}(z)V_{\alpha_1}(z_1)\dots V_{\alpha_n}(z_n)\rangle=0.
\end{equation}
When the number of external points $z_k$ equals $3$ this equation can be reduced to the ordinary Riemann differential equation (which is equivalent to the hypergeometric equation)
\begin{equation}\label{Phi12-4point-equation}
  \Biggl[\partial_z^2+
   b^2\Biggl\{\sum_{k=1}^3\Biggl(\frac{\Delta(\alpha_k)}{(z-z_k)^2}-\frac{1}{(z-z_k)}\partial_z\Biggr)-
  \sum_{i<j}\frac{\delta+\Delta_{ij}}{(z-z_i)(z-z_j)}\Biggr\}\Biggr]
  \langle V_{-\frac{b}{2}}(z)V_{\alpha_1}(z_1)V_{\alpha_2}(z_2)V_{\alpha_3}(z_3)\rangle=0,
\end{equation}
where
\begin{equation*}
   \delta=\Delta\left(-\frac{b}{2}\right)\qquad\text{and}\qquad
   \Delta_{12}=\Delta(\alpha_1)+\Delta(\alpha_2)-\Delta(\alpha_3)\qquad\text{etc}.
\end{equation*}
It means that the four-point correlation function with one degenerate field $V_{-\frac{b}{2}}$ can be expressed in terms of the hypergeometric functions\footnote{One has to remember that a differential equation similar to \eqref{Phi12-4point-equation} but with opposite chirality $z\rightarrow\bar{z}$ is also valid and the four-point correlation function is a bilinear combination $M_{ij}H_i(z)\bar{H}_j(\bar{z})$ where $H_i(z)$ and $\bar{H}_j(\bar{z})$ are  solutions of the corresponding differential equations.}. A more general four-point correlation function with one degenerate field $V_{-\frac{mb}{2}}$
\begin{equation}\label{4point-def}
  \langle V_{-\frac{mb}{2}}(z,\bar{z})V_{\alpha_1}(0)V_{\alpha_2}(1)V_{\alpha_3}(\infty)\rangle
\end{equation}
satisfies an ordinary differential equations of order $m+1$ in both variables $z$ and $\bar{z}$ as a consequence of the null-vector condition. A procedure to find differential operator which "kills" the correlation function \eqref{4point-def} from the null-vector condition is straitforward and was described in Ref. \cite{Belavin:1984vu}. However this procedure becomes tedious for large values of the parameter $m$ and it is difficult to construct such a differential operator explicitly using this method. Instead in Ref.s \cite{Fateev:2007tt,Fateev:2007qn} an explicit representation for the correlation function \eqref{4point-def} was obtained. Namely, the correlation function \eqref{4point-def} can be expressed in terms of a $2m$-dimensional Coulomb integral  
\begin{equation}\label{Liouville_4point}
  \begin{gathered}
  \langle V_{-\frac{mb}{2}}(x,\bar{x})V_{\alpha_1}(0)V_{\alpha_2}(1)V_{\alpha_3}(\infty)\rangle=
  \Omega_m(\alpha_1,\alpha_2,\alpha_3)\;\vert x\vert^{2mb\alpha_1}
  \vert x-1\vert^{2mb\alpha_2}\,\mathbf{J}_m(A,B,C|x),\\
  \mathbf{J}_m(A,B,C|x)=
  \int\hspace*{-5pt}...\hspace*{-5pt}\int\prod_{k=1}^m
  \vert t_k\vert^{2A}\vert t_k-1\vert^{2B}
  \vert t_k-x\vert^{2C}\prod_{i<j}|t_i-t_j|^{-4b^2}\,d^2t_1\dots d^2t_m
 \end{gathered}
\end{equation}
with parameters
\addtocounter{equation}{-1}
\begin{subequations}
\begin{equation}
       A=b\left(\alpha-2\alpha_1-Q+mb/2\right),\;\;
       B=b\left(\alpha-2\alpha_2-Q+mb/2\right),\;\;
       C=b\left(Q+mb/2-\alpha\right)
\end{equation}
and the normalization constant $\Omega_m(\alpha_1,\alpha_2,\alpha_3)$ is
\begin{equation}\label{O_m}
   \Omega_m(\alpha_1,\alpha_2,\alpha_3)=
   (-\pi\mu)^m
   \Bigl[\pi\mu\gamma(b^2)b^{2-2b^2}\Bigr]^{\frac{(Q-\alpha-mb/2)}{b}}\;
   \frac{\Upsilon'(-mb)\prod_{k=1}^3\Upsilon(2\alpha_k)}
    {\Upsilon(\alpha-Q-\frac{mb}{2})\prod_{k=1}^3\Upsilon(\alpha-2\alpha_k+\frac{mb}{2})},
\end{equation}
\end{subequations}
here $\alpha=\alpha_1+\alpha_2+\alpha_3$. Integration in \eqref{Liouville_4point} over each two-dimensional variable $t_k$ goes over the plane. This integral is convergent for some values of the parameters $\alpha_k$ and $b$, otherwise it should understood by means of analytic continuation.  The result \eqref{Liouville_4point} was obtained from arguments which did not involve the explicit construction of the differential equation. However, for several important purposes one needs the differential operator for the four-point correlation function in explicit form.

\begin{figure}
\psfrag{a1}{$0$}
\psfrag{a2}{$\frac{\pi}{2}$}
\psfrag{a3}{$\pi$}
\psfrag{a4}{$\frac{\pi\tau}{2}$}
\psfrag{a5}{$\pi\tau$}
\psfrag{a7}{$\frac{\pi+\pi\tau}{2}$}
\psfrag{b1}{$0$}
\psfrag{b2}{$1$}
\psfrag{b3}{$\infty$}
\psfrag{b4}{$x$}	  
        \centering
	\includegraphics[width=.9\textwidth]{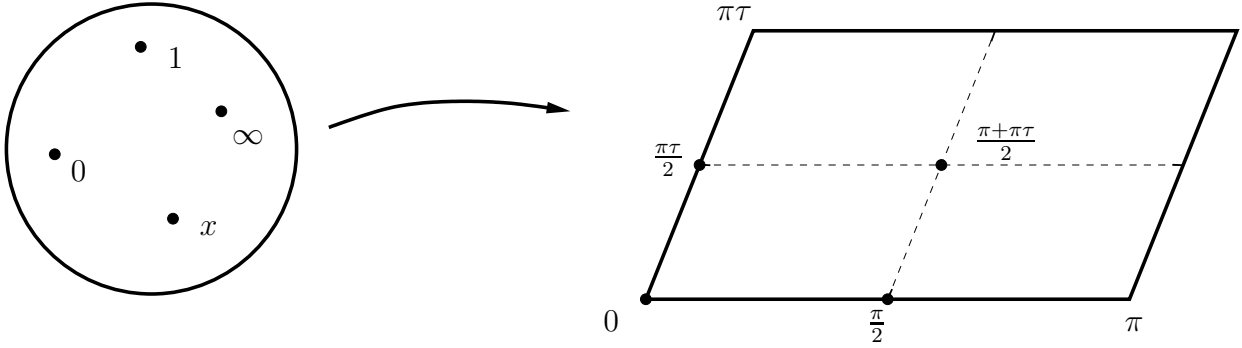}
	\caption{Transformation \eqref{u-map} maps a two-sheeted covering of the sphere with 
                 four marked points onto a torus which is 
                 represented by the parallelogram with the periods $\pi$ and $\pi\tau$. Points $x$, $0$, $1$ 
               and $\infty$ are mapped to the points $0$, 
               $\frac{\pi}{2}$, $\frac{\pi\tau}{2}$ and $\frac{\pi+\pi\tau}{2}$ on a torus.}
	\label{torus}
\end{figure}

In this paper we find the differential equation for the correlation function \eqref{4point-def}. Partially we follow the logic of paper \cite{Bauer:1991ai}. Let us consider the five-point correlation function with one degenerate field
\begin{equation}
\langle V_{-\frac{1}{2b}}(z)V_{\alpha_1}(0)V_{\alpha_2}(1)V_{\alpha_3}(\infty)V_{\alpha_4}(x)\rangle.
\end{equation}
This correlation function satisfies a second order partial differential equation. For future purposes it is convenient to define the function $\Psi(u|q)$ as
\begin{equation}\label{5-point}
  \langle V_{-\frac{1}{2b}}(z)V_{\alpha_1}(0)V_{\alpha_2}(1)V_{\alpha_3}(\infty)V_{\alpha_4}(x)\rangle=
  z^{\frac{1}{2b^2}}(z-1)^{\frac{1}{2b^2}}
  \frac{\left(z(z-1)(z-x)\right)^{\frac{1}{4}}}
  {\left(x(x-1)\right)^{\frac{2\Delta(\alpha_4)}{3}+\frac{1}{12}}}
  \frac{\Theta_1(u)^{b^{-2}}}{\Theta'_1(0)^{\frac{b^{-2}+1}{3}}}\,\Psi(u|q),
\end{equation}
where the variable $u$ is related with the variables $z$ and $x$ as
\begin{equation}\label{u-map}
    u=\frac{\pi}{4K(x)}\int_0^{\frac{z-x}{x(z-1)}}\frac{dt}{\sqrt{t(1-t)(1-xt)}},
\end{equation}
and $\Theta_1(u)$ is the Jacobi theta function (see definition in appendix \ref{formulae}). We note that \eqref{u-map} maps a two-sheeted covering of the sphere with four marked points $x$, $0$, $1$ and $\infty$ onto a torus $T$ with periods $\pi$ and $\pi\tau$ where $\tau$ is given by \eqref{tau}. Marked points $x$, $0$, $1$ and $\infty$ are mapped to the points $0$, $\frac{\pi}{2}$, $\frac{\pi\tau}{2}$ and  $\frac{\pi+\pi\tau}{2}$ on a torus (see fig. \ref{torus}).
The function $\Psi(u|q)$  defined by eq \eqref{5-point} satisfies the non-stationary Schr\"odinger equation with doubly periodic potential
\begin{equation}\label{2order-diff-2}
   \Biggl[\partial^2_u-\mathbb{U}(u)+\frac{4i}{\pi b^2}\partial_{\tau}\Biggr]\Psi(u|q)=0,
\end{equation}
where the potential $\mathbb{U}(u)$ is given by
\begin{equation}\label{Potential-U}
   \mathbb{U}(u)=\mathbb{V}(u)+\delta\mathbb{V}
\end{equation}
with 
\begin{equation}\label{double-periodic-potential}
  \mathbb{V}(u)=\sum_{j=1}^4s_j(s_j+1)\wp(u-\omega_j)
\end{equation}
and parameters $s_k$ are related with parameters $\alpha_k$ as
\begin{equation}\label{ak-sk}
    \alpha_k=\frac{Q}{2}-\frac{b}{2}\left(s_k+\frac{1}{2}\right).
\end{equation}
In \eqref{double-periodic-potential}  $\wp(u)$ is the Weierstra\ss  ~elliptic function with periods $\pi$ and $\pi\tau$ defined by the infinite sum
\begin{equation}\label{Wp}
  \wp(u)=\frac{1}{u^2}+\sum_{k^2+l^2\neq0}\left(\frac{1}{(u-\pi k-\pi\tau l)^2}-
  \frac{1}{(\pi k+\pi\tau l)^2}\right),
\end{equation}
and $\omega_j$ are the half periods
\begin{equation}\label{half-periods}
   \omega_1=\frac{\pi}{2},\qquad\omega_2=\frac{\pi\tau}{2},\qquad
   \omega_3=\omega_1+\omega_2,\qquad\omega_4=0.
\end{equation}
We choose the $u$-independent term $\delta\mathbb{V}$ in \eqref{Potential-U} in such a way that the expansion of the potential $\mathbb{U}(u|x)$ at the origin has the form\footnote{This $u$-independent term is $\delta\mathbb{V}=-\sum_{k=1}^3s_k(s_k+1)\wp(\omega_k)$.}
\begin{equation}
  \mathbb{U}(u|x)=\frac{s_4(s_4+1)}{u^2}-
   4\sum_{k=1}^{\infty}\frac{W^{(k+1)}(\tau)}{k!^2}u^{2k},
\end{equation}
where
\addtocounter{equation}{-1}
\begin{subequations}
\begin{equation}\label{Current-general}
  W^{(k)}(\tau(x))=\left(\frac{2K(x)}{\pi}\right)^{2k}
   \Bigl((-1)^{k+1}(x-1)\mathsf{w}^{(k)}_1P_k(x)+x\mathsf{w}^{(k)}_2P_k(1-x)
  -(-1)^{k-1}x(x-1)\mathsf{w}^{(k)}_3x^{k-2}P_k(1/x)\Bigr)
\end{equation}
with
\begin{equation}\label{w-k}
  \mathsf{w}^{(k)}_j=\Biggl[s_j(s_j+1)+
    \frac{s_4(s_4+1)}{(4^k-1)}\Biggr].
\end{equation}
\end{subequations}
In eq \eqref{Current-general} $P_k(x)$ are polynomials of degree $k-2$
\begin{equation}\label{polynomials}
  \begin{aligned}
    &P_2(1-x)=1,\\
    &P_3(1-x)=-\frac{4}{3}\,(1+x),\\
    &P_4(1-x)=\frac{4}{5}\,(2+13x+2x^2),\\
    &P_5(1-x)=-\frac{64}{35}\,(1+30x+30x^2+x^3),\\
    &\qquad\dots\dots\dots\dots\dots\dots\dots\dots\dots\dots\dots\dots\dots
  \end{aligned}
\end{equation}
These polynomials can be obtained as an expansion of the square of the elliptic sine (see appendix \ref{formulae} for connection between the Weierstra\ss ~function and the elliptic sine function). Namely
\begin{equation}\label{Polynomials-generating-function}
   \sn^2(t|\sqrt{x})=\sum_{k=1}^{\infty}\frac{P_{k+1}(1-x)}{k!^2}\,t^{2k}.
\end{equation}
The polynomials $P_k(1-x)$ defined by eq \eqref{polynomials} can be rapidly calculated using the recursive formula
\begin{equation}
   m(2m-1)P_{m+1}(1-x)=3x\sum_{k=1}^{m}\frac{m!^2}{(k-1)!^2(m-k)!^2}P_k(1-x)P_{m+1-k}(1-x)
\end{equation}
with initial condition
\begin{equation*}
   P_1(1-x)=-\frac{1}{3}\left(1+\frac{1}{x}\right);\qquad P_2(1-x)=1.
\end{equation*}
It is important also to have in mind the expansion
\begin{equation}
   \sn^{-2}(t|\sqrt{x})=\frac{1}{t^2}+\frac{1}{3}(x+1)+
   \sum_{k=1}^{\infty}\frac{G_{k+1}(1-x)}{k!^2}\,t^{2k},
\end{equation}
where
\begin{equation*}
  G_k(x)=\frac{1}{4^k-1}\Bigl((-1)^{k-1}(x-1)P_k(x)+xP_k(1-x)-(-1)^{k-1}x(x-1)x^{k-2}P_k(1/x)\Bigr).
\end{equation*}

Let us try to find a solution to the differential equation \eqref{2order-diff-2} in the form\footnote{As we will see later the function $\Psi(\tau)$ defines the four-point correlation function of the fields $V_{\alpha_k}$ with $k=1,2,3$ and the field $V_{\alpha_4-\frac{1}{2b}}$. The functions $\Psi_{-k}(\tau)$ define the four-point correlation function of the fields $V_{\alpha_k}$ and  some descendants of the field $V_{\alpha_4-\frac{1}{2b}}$.}
\begin{equation}\label{Psi-anzatz}
  \Psi(u|q)= 
  u^{s_4+1}\left(\Psi(\tau)+\Psi_{-1}(\tau)u^2+\Psi_{-2}(\tau)u^4+\dots\right),
\end{equation}
then \eqref{2order-diff-2} will be equivalent to the semi-infinite matrix differential equation \cite{Bauer:1991ai}
\begin{equation}\label{WZ1}
   \Biggl(-J_{-}+\frac{i}{\pi b^2}\frac{\partial}{\partial\tau}+
   \sum_{k=1}^{\infty}\frac{W^{(k+1)}(\tau)}{k!^2}J_{+}^k\Biggr)\vec{\Psi}(\tau)=0,
\end{equation}
where\footnote{The general term below the diagonal in $J_{-}$ in \eqref{Js} is equal to $-n(s_4+n+\frac{1}{2})$.}
\begin{equation}\label{Js}
  J_{-}= 
  \begin{pmatrix}
       \dots&\dots&\dots&\dots\\
       \dots&0&0&0\\
       \dots&-2s_4-5&0&0\\
       \dots&0&-s_4-\frac{3}{2}&0
   \end{pmatrix}\quad
  J_{+}= 
  \begin{pmatrix}
       \dots&\dots&\dots&\dots\\
       \dots&0&1&0\\
       \dots&0&0&1\\
       \dots&0&0&0
   \end{pmatrix}\quad
     \vec{\Psi}(\tau)=
   \begin{pmatrix}
       \dots\\
       \Psi_{-2}(\tau)\\
       \Psi_{-1}(\tau)\\
       \Psi(\tau)
   \end{pmatrix}
\end{equation}
Or explicitly
\begin{equation}\label{WZ2}
  \begin{aligned}
    &(s_4+\frac{3}{2})\Psi_{-1}+\frac{i}{\pi b^2}\partial_{\tau}\Psi=0,\\
    &(2s_4+5)\Psi_{-2}+\frac{i}{\pi b^2}\partial_{\tau}\Psi_{-1}+W^{(2)}(\tau)\Psi=0,\\
    &\dots\dots\dots\dots\dots\dots\dots\dots\dots\dots\dots\dots\dots
  \end{aligned}
\end{equation}
We see that if the parameter $s_4$ in eq \eqref{ak-sk} takes the values\footnote{It corresponds to the situation $\alpha_4=\frac{1}{2b}-\frac{mb}{2}$ and hence in the operator product expansion  $V_{-\frac{1}{2b}}(z)V_{\alpha_4}(x)$ appears the degenerate field $V_{-\frac{mb}{2}}$.} 
\begin{equation}
  s_4=-m-\frac{3}{2}
\end{equation}
then the infinite chain of equations \eqref{WZ1} has a finite sub-chain which can be written in the form \eqref{WZ1}, but with finite $(m+1)\times(m+1)$  matrices $J_{+}$ and $J_{-}$\footnote{More precisely, in this case we meet the situation when one of the solutions to the differential equation \eqref{2order-diff-2} contains logarithmic terms in the variable $u$ (the so-called resonance case). The condition that these logarithmic terms are cancelled is equivalent to the condition that the finite-dimensional subchain of differential equations  \eqref{WZ1} with matrices $J_{+}$ and $J_{-}$ given by eqs \eqref{Js-finite} is satisfied.}:
\begin{equation}\label{Js-finite}
  J_{-}= 
  \begin{pmatrix}
       0&0&\dots&0&0&0\\
       m&0&\dots&0&0&0\\
       \dots&\dots&\dots&\dots&\dots\\
       0&0&\dots&2(m-1)&0&0\\
       0&0&\dots&0&m&0
   \end{pmatrix},\qquad
  J_{+}= 
  \begin{pmatrix}
       0&1&\dots&0&0\\
       0&0&\dots&0&0\\
       \dots&\dots&\dots&\dots&\dots\\
       0&0&\dots&0&1\\
       0&0&\dots&0&0
   \end{pmatrix}.
\end{equation}
From this finite chain of equations we can conclude that the function $\Psi(\tau)$ in eq \eqref{Psi-anzatz} satisfies a differential equation of the order $(m+1)$. This equation can be written in the form\footnote{Here for convenience we rescaled $W^{(k)}(\tau)\rightarrow\left(\frac{\pi b^2}{i}\right)^{k}W^{(k)}(\tau)$.}
\begin{equation}\label{DIFF-n}
  \Bigl[\partial_{\tau}^{m+1}+\frac{m(m+1)(m+2)}{6}W^{(2)}(\tau)\partial_{\tau}^{m-1}+
  \dots\Bigr]\Psi(\tau)=0.
\end{equation}
All functions $W^{(k)}(\tau)$ will enter in the differential equation \eqref{DIFF-n} with positive integer coefficients.  It is reasonable to change variable $\tau\rightarrow x$ in \eqref{DIFF-n} using projective invariance. Namely, the differential equation \eqref{DIFF-n} is invariant under the change of variables
\begin{equation}\label{invariance}
  \begin{gathered}
   \tau\rightarrow\omega(\tau),\quad\Psi(\tau)\rightarrow\Bigl(\frac{d\omega}{d\tau}\Bigr)^{-\frac{m}{2}}
  \Psi(\omega),\\
   W^{(2)}(\tau)\rightarrow\Bigl(\frac{d\omega}{d\tau}\Bigr)^{2}W^{(2)}(\omega)+
   \frac{1}{2}\{\omega,\tau\},\qquad
   W^{(k)}(\tau)\rightarrow\Bigl(\frac{d\omega}{d\tau}\Bigr)^{k}W^{(k)}(\omega)
   \quad\text{for}\quad k>2,
  \end{gathered}
\end{equation}
where $\{f,z\}$ is the Schwartz derivative, which is defined as $\{f,z\}=f'''/f'-3/2(f''/f')^2$. Choosing
\begin{equation}
   \tau=i\frac{K(1-x)}{K(x)},\qquad\frac{d\tau}{dx}=\frac{i\pi}{4x(x-1)K^2(x)}
\end{equation}
and using that in this case
\begin{equation}
  \{x,\tau\}=\frac{1}{2x^2(1-x)}+\frac{1}{2(x-1)^2x}-\frac{1}{2x(1-x)}
\end{equation}
we obtain that the function $\Psi(x)$  satisfies the same differential equation which can be easily derived from the system \eqref{WZ1} with matrices $J_{+}$ and $J_{-}$ given by \eqref{Js-finite} and the substitution $\frac{i}{\pi b^2}\frac{\partial}{\partial\tau}\rightarrow\frac{\partial}{\partial x}$
\begin{equation}\label{DIFF-n-x}
 \Bigl[\partial_{x}^{m+1}+\frac{m(m+1)(m+2)}{6}W^{(2)}(x)\partial_{x}^{m-1}+\dots\Bigr]\Psi(x)=0,
\end{equation}
but with currents $W^{(k)}(x)$ given by\footnote{Here we are going back and use the parameters $\Delta_k=\alpha_k(Q-\alpha_k)$ instead of the parameters $s_k$ defined by \eqref{ak-sk}.}\addtocounter{equation}{-1}
\begin{subequations}
\begin{equation}\label{currents-DIFF-n-x}
 W^{(k)}(x)=\frac{\mathsf{w}^{(k)}_1P_k(x)}{x^k(1-x)^{k-1}}+
 \frac{\mathsf{w}^{(k)}_2P_k(1-x)}{(x-1)^kx^{k-1}}-
 \frac{\mathsf{w}^{(k)}_3x^{k-2}P_k(1/x)}{x^{k-1}(1-x)^{k-1}},
\end{equation}
where
\begin{equation}\label{quantum-numbers-DIFF-n-x}
  \begin{aligned}
    &\mathsf{w}^{(2)}_j=\Biggl[
    b^{2}\Bigl(\Delta_j-\frac{1}{2}\Bigr)-
    \frac{b^{4}}{60}\bigl((m+1)^2+11\bigr)\Biggr],\\
    &\mathsf{w}^{(k)}_j=\Biggl[
    b^{2k-2}\Bigl(\Delta_j-\frac{1}{2}-\frac{1}{4b^2}\Bigr)-
    \frac{b^{2k}}{4S_k}\bigl((m+1)^2+Y_k\bigr)\Biggr]\;\text{if}\;\;k>2
  \end{aligned}
\end{equation}
with
\begin{equation}
  S_k=4^k-1;\;\;\;\;\;\;\;\;Y_k=3\cdot4^{k-1}-1
\end{equation}
\end{subequations}
and polynomials $P_k(1-x)$ defined by eqs \eqref{polynomials}, \eqref{Polynomials-generating-function}. We note that the function $\Psi(x)$ in \eqref{DIFF-n-x} is related with the four-point correlation function with one degenerate field $V_{-\frac{mb}{2}}(x)$ as (we write down only the chiral part)
\begin{equation}
   \Psi(x)=x^{-\frac{m(m+2)}{6}b^2}(x-1)^{-\frac{m(m+2)}{6}b^2}
   \langle V_{-\frac{mb}{2}}(x)V_{\alpha_1}(0)V_{\alpha_2}(1)V_{\alpha_3}(\infty)\rangle
\end{equation}
We have checked the first seven differential operators obtained by the above method with those obtained by the "brute-force method" and found a complete agreement. We list the first few examples of the differential equations in the appendix \ref{Examples}. We see that the coefficients before all terms in the differential operator \eqref{DIFF-n-x} are positive integer numbers which can be easily obtained from the matrix equation \eqref{WZ1} with matices $J_{+}$ and $J_{-}$ given by \eqref{Js-finite} and have a combinatorial structure.

It is instructive to notice that the coefficients $W^{(k)}(x)$ in the differential operator \eqref{DIFF-n-x} transform as
\begin{equation}\label{Currents-duality}
   W^{(k)}(x)\rightarrow(-1)^kW^{(k)}(x),
\end{equation}
under the substitution $b\rightarrow ib$, $\Delta_k\rightarrow1-\Delta_k$, which corresponds to the differential operator killing the four-point correlation function with one degenerate field in the complementary theory with central charge $c=26-c_L$ and with additional fields with conformal dimensions $\tilde{\Delta}_k=1-\Delta_k$. Transformation \eqref{Currents-duality} reflects a very simple property of the differential operator of type \eqref{DIFF-n-x} with arbitrary functions (currents) $W^{(k)}(x)$ \cite{Bilal:1988jf}. Namely, if $\psi_1,\dots,\psi_{m+1}$ are some $(m+1)$ linearly independent solutions to \eqref{DIFF-n-x} then $\chi_1,\dots,\chi_{m+1}$, where $\chi_k$ is given by the Wronskian
\begin{equation}
    \chi_k=\begin{vmatrix}
    \psi_{1}  & \dots& \psi_{k-1}& \psi_{k+1}  & \dots & \psi_{m+1}\\
    \partial\psi_{1} &\dots& \partial\psi_{k-1}& \partial\psi_{k+1} & \dots & \partial\psi_{m+1}\\
    \dots&\hdotsfor{4}&\dots\\
    \dots&\hdotsfor{4}&\dots\\
    \partial^{m-1}\psi_{1}&\dots &\partial^{m-1}\psi_{k-1}&\partial^{m-1}\psi_{k+1} 
    &\dots & \partial^{m-1}\psi_{m+1} 
   \end{vmatrix}
\end{equation}
are solutions to the differential equation \eqref{DIFF-n-x} with currents $W^{(k)}(x)\rightarrow(-1)^kW^{(k)}(x)$. It means that the corresponding correlation functions in both theories are simply related. Another interesting property of the differential operator \eqref{DIFF-n-x} under the transformation \eqref{Currents-duality} is that it transforms up to the sign $(-1)^{m+1}$ to the hermitian conjugate operator.

Let us mention also that we obtain an important example of a differential operator with a unitarized monodromy matrix. In general, if we consider the differential equation \eqref{DIFF-n-x} with currents given by \eqref{currents-DIFF-n-x} but now with arbitrary numbers $\mathsf{w}^{(k)}_j$ and polynomials $P_k(x)$ of degree $(k-2)$, then it gives the most general Fuchsian differential equation with three singular points $0$, $1$ and $\infty$ with singularities defined by $\mathsf{w}^{(k)}_j$\footnote{By gauge transformation $\Psi(x)\rightarrow x^{\alpha}(x-1)^{\beta}\Psi(x)$ one can always set the coefficient before the term with subleading derivative to zero.}. Without loss of generality we can assume that the polynomials $P_k(x)$ are normalized as
\begin{equation}\label{general-polynomial-normalization}
   P_k(x)=1+\dots
\end{equation}
The subleading coefficients in polynomials \eqref{general-polynomial-normalization} define so called accessory parameters which do not affect on singular behavior but of course contribute to the monodromy. The problem is to tune them in such a way that monodromy is unitarized. With a requirement of that type we deal with the solution of the $\mathfrak{sl}[n]$ Toda on a sphere with three singular points \cite{Fateev:2007ab}. The unitary monodromy allows one to build a bilinear combination of holomorphic and antiholomorphic solutions which is a single-valued function on a sphere with three punctures. In paper \cite{Fateev:2007ab} it was proposed that the solution to this problem is unique in some special domain of the parameters $\mathsf{w}^{(k)}_j$. It means that all coefficients in \eqref{general-polynomial-normalization} i.e. accessory parameters are fixed. However this statement is not yet proved (at least to our knowledge) nevertheless it looks reasonable from a physical point of view. The fixing of the accessory parameters is a very transcendental problem and its solution is far from being done, but in some cases one can guess an answer. Here we have found a class of differential equations with unitarized monodromy which is parameterized by four numbers $\Delta_1$, $\Delta_2$, $\Delta_3$ and $b$ and the accessory parameters or what is the same the polynomials $P_k(x)$ given by \eqref{polynomials}. It follows trivially from the fact that in this case the solution to the corresponding differential equation represents the holomorphic part of the four-point correlation function \eqref{Liouville_4point} which is obviously a single-valued function because it is given by a multiple integral over the plane. This class of differential operators can be a starting point for the numerical study of the accessory parameters problem.

All results of this section can be trivially rewritten for the correlation function with degenerate  field $V_{-\frac{m}{2b}}$, but for general degenerate fields $V_{\alpha_{mn}}$ with $m\neq0$ and $n\neq0$ this construction is more complicated and we will not discuss it here (see also section \ref{Conclusion}).   
\section{Integrable potentials and conformal blocks}\label{Int-potentials}
In the previous section we constructed the differential operator for the four-point correlation function \eqref{4point-def} which has an explicit integral representation. In this section we consider the differential equation \eqref{2order-diff-2} and for special choice of the parameters $s_k$ find an explicit integral representation for its solution.

By the transformation
\begin{equation}
  \Psi(u|q)\rightarrow x^{\lambda_1}(x-1)^{\lambda_2}\Psi(u|q)
\end{equation}
where
\begin{equation*}
  \begin{aligned}
    &\lambda_1=\frac{1}{3}\left(2\left(\alpha_1-Q/2\right)^2-\left(\alpha_2-Q/2\right)^2
     -\left(\alpha_3-Q/2\right)^2\right),\\
    &\lambda_2=\frac{1}{3}\left(2\left(\alpha_2-Q/2\right)^2-\left(\alpha_1-Q/2\right)^2
     -\left(\alpha_3-Q/2\right)^2\right),
 \end{aligned}
\end{equation*}
the differential equation \eqref{2order-diff-2} can be transformed to the generalized Lam\'e heat equation
\begin{equation}\label{Diff-double-periodic}
   \left(-\partial^2_u+\mathbb{V}(u)\right)\Psi(u|q)=\frac{4i}{\pi b^2}\partial_{\tau}\Psi(u|q),
\end{equation}
where the potential $\mathbb{V}(u)$ is given by \eqref{double-periodic-potential}. For integer $s_j$ the potential \eqref{double-periodic-potential} is known to be a finite-gap potential and called the Treibich-Verdier potential \cite{Treibich}. The differential equation \eqref{Diff-double-periodic} was studied in particular cases in Ref.s \cite{Etingof:1993gk,felder-2002,Felder:2000mq} (see also Ref. \cite{Bazhanov:2004hv} where interesting relation with Baxter's $Q-$operator for the eight-vertex model was discovered). We propose, that if all parameters $s_k$ in eqs \eqref{double-periodic-potential}-\eqref{ak-sk} take the values
\begin{equation}\label{s_k}
  s_k=m_k+\frac{2n_k}{b^2},
\end{equation}
then the differential equation \eqref{Diff-double-periodic} is integrable in the sense that the general solution to \eqref{Diff-double-periodic} can obtained from the general solution to the heat equation:
\begin{equation}\label{Diff-double-periodic-free}
  \partial^2_u\Psi(u|q)+\frac{4i}{\pi b^2}\partial_{\tau}\Psi(u|q)=0
\end{equation}
by an appropriate integral transformation. For example let $s_1=s_2=s_3=0$ and $s_4=1$, then the general solution to the equation
\begin{equation}\label{Diff-double-periodic-1}
   \left(-\partial^2_u+2\wp(u)\right)\Psi(u|q)=\frac{4i}{\pi b^2}\partial_{\tau}\Psi(u|q),
\end{equation}
can be obtained from the general solution $\Psi_0(u|q)$ of the  heat equation \eqref{Diff-double-periodic-free} as follows:
\begin{equation}\label{Lame-Sol-1}
  \Psi(u|q)=\int_0^{\pi}\left(\frac{\Theta_1(v)}{\Theta'_1(0)^{\frac{1}{3}}}\right)^{b^2}
  \frac{E(u+v)}{E(u)E(v)}\,\Psi_0(u+b^2v|q)\,dv,  
\end{equation}
where we introduced the notation
\begin{equation}
    E(u)=\frac{\Theta_1(u)}{\Theta_1'(0)}.
\end{equation}
The proof of \eqref{Lame-Sol-1} and of more general relations can be found in the appendix \ref{Andre-proof}. In principle, the integration in \eqref{Lame-Sol-1} can go between any two zeroes of $\Theta_1(v)$, for example between $0$ and $\pi$ or between $0$ and $\pi\tau$. It is instructive to consider the limit $b\rightarrow\infty$ and take in \eqref{Lame-Sol-1}
\begin{equation}
    \Psi_0(u+b^2v|q)=q^{\frac{b^2\lambda^2}{4}}e^{-\lambda(u+b^2v)},    
\end{equation}
with $\lambda$ being fixed at $b\rightarrow\infty$. In this case the integral in \eqref{Lame-Sol-1} is governed by the saddle point $\nu$, which is solution to the equation
\begin{equation}
     \frac{\Theta_1'(\nu)}{\Theta_1(\nu)}=\lambda
\end{equation}
and $\Psi(u|q)$ has a limit (up to irrelevant factors):
\begin{equation}\label{Stat-Lame-solution}
     \Psi(u|q)\rightarrow\frac{\Theta_1(u+\nu)}{\Theta_1(u)}e^{-\lambda u},
\end{equation}
which is the solution to the stationary Lam\'e equation with energy $\wp(\nu)$. So, the solution \eqref{Lame-Sol-1} can be viewed as a "quantization" of the solution to the stationary equation \eqref{Stat-Lame-solution}. More suprising is that equation \eqref{Diff-double-periodic} is integrable for values of the parameters $s_k$ which vanish in the limit $b\rightarrow\infty$. Let us consider for example the case $s_1=s_2=s_3=0$  and $s_4=\frac{2}{b^2}$, then the general solution to
\begin{equation}\label{Diff-double-periodic-2}
   \left(-\partial^2_u+\frac{2}{b^2}\left(\frac{2}{b^2}+1\right)
   \wp(u)\right)\Psi(u|q)=\frac{4i}{\pi b^2}\partial_{\tau}\Psi(u|q)
\end{equation}
is given by the integral transform
\begin{equation}\label{Lame-Sol-2}
  \Psi(u|q)=\Theta'_1(0)^{\frac{2}{3}(1-\frac{2}{b^2})}
  \int_0^{\pi}\left(\frac{\Theta_1(v)}{\Theta'_1(0)^{\frac{1}{3}}}\right)^{\frac{4}{b^2}}
  \left(\frac{E(u+v)}{E(u)E(v)}\right)^{\frac{2}{b^2}}\,\Psi_0(u+2v|q)\,dv,
\end{equation}
where $\Psi_0(u|q)$ is the solution to \eqref{Diff-double-periodic-free}. In the general case \eqref{s_k} solution is likely to be  given by an integral of dimension 
\begin{equation}\label{N}
  N=g+n_1+n_2+n_3+n_4,
\end{equation} 
where $g$ is the number of gaps for the classical potential \cite{MR1339715}
\begin{equation}\label{g}
    g=\frac{1}{2}\left(2\max m_k,1+\mathtt{m}-(1+(-1)^{\mathtt{m}})\bigl(\min m_k+\frac{1}{2}\bigr)\right),
\end{equation}
here $\mathtt{m}=\sum m_k$. We justified this hypothesis in many cases where an explicit solution has been constructed. We believe that this conjecture is true, however for general integer values of $m_k$ and $n_k$ an explicit integral representation for the solution at present is not known. In many particular cases four-point conformal blocks (see below) which can be viewed as a limiting value of the solution to \eqref{Diff-double-periodic}  are given by the integrals of elliptic functions. Some of them  are listed in appendix \ref{Conf-blocks}.

Although we assume that equation \eqref{Diff-double-periodic} is integrable for all values of the parameters $s_k$ given by \eqref{s_k} the solution can be very non-trivial. All explicit examples known to us show that the solution depends irregularly on the parameters $s_k$ and it is difficult  to obtain closed expression for all values \eqref{s_k}. Below we consider the case $s_1=s_2=s_3=0$ and $s_4=m$, which corresponds to the five point  correlation function  
\begin{multline}\label{special-4point-definition}
   \langle V_{-\frac{1}{2b}}(z,\bar{z})
   V_{\frac{1}{2b}-\frac{(2m-1)b}{4}}(x,\bar{x})V_{\eta}(0)V_{\eta}(1)V_{\eta}(\infty)\rangle=\\=
   \left|\frac{z(z-1)}{x(x-1)}\right|^{\frac{1}{2}+\frac{1}{b^2}}\left|z-x\right|^{\frac{1}{2}}
   \left|x(x-1)\right|^{\frac{b^2}{3}(m+\frac{1}{2})^2-\frac{(b^2-1)(b^2+2)}{3b^2}}
   \frac{\left|\Theta_1(u|q)\right|^{\frac{2}{b^2}}}
   {\left|\Theta'_1(0|q)\right|^{\frac{2}{3b^2}+\frac{2}{3}}}
   \boldsymbol{\Psi_m}(u,\bar{u}),
\end{multline}
where
\begin{equation}
   \eta=\frac{Q}{2}-\frac{b}{4}.
\end{equation}
The function $\boldsymbol{\Psi_m}(u,\bar{u})$ satisfies the differential equation\footnote{It satisfies also the antiholomorphic differential equation similar to \eqref{Diff-double-periodic-3} but with $u\rightarrow\bar{u}$ and $\tau\rightarrow\bar{\tau}$.}
\begin{equation}\label{Diff-double-periodic-3}
   \left(-\partial^2_u+m(m+1)\wp(u)\right)\boldsymbol{\Psi_m}(u,\bar{u})
   =\frac{4i}{\pi b^2}\partial_{\tau}\boldsymbol{\Psi_m}(u,\bar{u}),
\end{equation}
whose general solution is given by the $m$-dimensional integral (see appendix \ref{Andre-proof})
\begin{equation}\label{Lame-sol-m}
   \Psi(u|q)=
   \int\limits_{0}^{\pi}\hspace*{-5pt}...\hspace*{-5pt}\int\limits_{0}^{\pi}
   \prod_{k=1}^m\left(\frac{\Theta_1(v_k)}{\Theta_1'(0)^{\frac{1}{3}}}\right)^{mb^2}
   \prod_{i<j}\left|\frac{\Theta_1(v_i-v_j)}{\Theta_1'(0)^{\frac{1}{3}}}\right|^{-b^2}
   \prod_{k=1}^m\frac{E(u+v_k)}{E(u)E(v_k)}\,\Psi_0(u+b^2v|q)\,
   dv_1...dv_m,
\end{equation}
where $v=v_1+\dots+v_m$ and $\Psi_0(u|q)$ is again some solution to the heat equation \eqref{Diff-double-periodic-free}.  Apparently, the space of solutions to \eqref{Diff-double-periodic-3} is infinite dimensional and spanned by the integrals \eqref{Lame-sol-m} with $\Psi_0(u|q)$ taken to be equal to
\begin{equation}\label{ploskaya-volna}
     \Psi_P^{\pm}(u|q)=q^{P^2}e^{\pm2b^{-1}Pu}.
\end{equation}
Each solution $\Psi_P^{\pm}(u|q)$ corresponds to a particular five-point conformal block. In this paper we consider four-point conformal blocks which can be obtained from the five-point conformal blocks in the limit $u\rightarrow0$. More precisely, there are two types of four-point conformal blocks which can be obtained from the solution \eqref{Lame-sol-m} in this limit. This is related with the fact that in the operator product of the degenerate field $V_{-\frac{1}{2b}}$ with any other fields in the correlation function \eqref{special-4point-definition} there appears only two primary fields together with their descendants. In particular, in the operator product of the field $V_{-\frac{1}{2b}}$ with the field $V_{\frac{1}{2b}-\frac{(2m-1)b}{4}}$ we obtain two terms (we used the reflection relation $\alpha\rightarrow Q-\alpha$ in the second term)
\begin{equation}\label{OPE-1/2b}
   V_{-\frac{1}{2b}}V_{\frac{1}{2b}-\frac{(2m-1)b}{4}}=
   \left[V_{-\frac{(2m-1)b}{4}}\right]+\left[V_{\frac{(2m+3)b}{4}}\right].
\end{equation}
The first term in the r.h.s. of \eqref{OPE-1/2b} corresponds to the main asymptotic of the solution \eqref{Lame-sol-m} at $u\rightarrow0$ which is $\Psi(u|q)\sim u^{-m}$ and gives the conformal block
\begin{equation}\label{Conf-Block-1}
  \mathcal{H}_m^{(P)}(q)\overset{\text{def}}{=}
  \mathfrak{H}_P\Bigl(\genfrac{}{}{0pt}{}{\frac{Q}{2}-\frac{b}{4}\;\;\;\;\;\;\frac{Q}{2}-\frac{b}{4}}
  {-\frac{(2m-1)b}{4}\;\;\frac{Q}{2}-\frac{b}{4}}\Bigl|q\Bigr),
\end{equation}
where $\mathfrak{H}_P(\dots|q)$ is defined by \eqref{Elliptic-Block-definition}.
Fortunately, in order to obtain the conformal block \eqref{Conf-Block-1} one has to take instead of $\Psi_0$ in \eqref{Lame-sol-m} one of the simplest solutions \eqref{ploskaya-volna} (it does not matter which one, because both of hem have the same asymptotic at $u\rightarrow0$).   
Having in mind definition \eqref{special-4point-definition} of the function $\boldsymbol{\Psi_m}(u,\bar{u})$, we obtain the integral representation for the conformal block \eqref{Conf-Block-1}
\begin{equation}\label{Elliptic-block}
  \mathcal{H}_m^{(P)}(q)=N^{-1}_m
  \int\limits_{0}^{\pi}\hspace*{-5pt}...\hspace*{-5pt}\int\limits_{0}^{\pi}e^{2bP(u_1+\dots+u_m)}
  \prod_{k=1}^{m}E(u_k)^{mb^2}\prod_{i<j}|E(u_i-u_j)|^{-b^2}\,du_1\dots du_m
\end{equation}
with normalization constant $N_m$ given by\footnote{Normalization constant \eqref{N-m}  and more general normalization constants were calculated in the appendix \ref{Fateev-integrals}. } 
\addtocounter{equation}{-1}
\begin{subequations}
\begin{equation}\label{N-m}
   N_m=\frac{\pi^m e^{\pi mbP}}{2^{\frac{m(m+1)b^2}{2}}}
   \frac{\prod_{k=1}^m\Gamma\left(1-\frac{kb^2}{2}\right)\Gamma\left(1+\frac{(2m+1-k)b^2}{2}\right)}
   {\Gamma^m\left(1-\frac{b^2}{2}\right)\prod_{k=1}^m\Gamma\left(1+\frac{kb^2}{2}+ibP\right)
   \Gamma\left(1+\frac{kb^2}{2}-ibP\right)}.
\end{equation}
\end{subequations}
We have performed an expansion of the integral \eqref{Elliptic-block} in series at $q\rightarrow0$ and compared it with a known expansion following  from  Alyosha Zamolodchikov's recursion formula for the conformal block \cite{Zamolodchikov:1985ie} and find complete agreement up to high orders.  

\begin{figure}
\psfrag{p}{$P$}\psfrag{c}{$\mathcal{C}$}
\psfrag{a1}{$\frac{ib}{2}$}\psfrag{b1}{$\scriptscriptstyle-\textstyle\frac{ib}{2}$}
\psfrag{a1m}{$\frac{imb}{2}$}\psfrag{b1m}{$\scriptscriptstyle-\textstyle\frac{imb}{2}$}
\psfrag{a2}{$\frac{ib}{2}+ib^{-1}$}\psfrag{b2}{$\scriptscriptstyle-\textstyle\frac{ib}{2}+ib^{-1}$}	  
\psfrag{a2m}{$\frac{imb}{2}+ib^{-1}$}\psfrag{b2m}{$\scriptscriptstyle-\textstyle\frac{imb}{2}+ib^{-1}$}	  
        \centering
	\includegraphics[width=.8\textwidth]{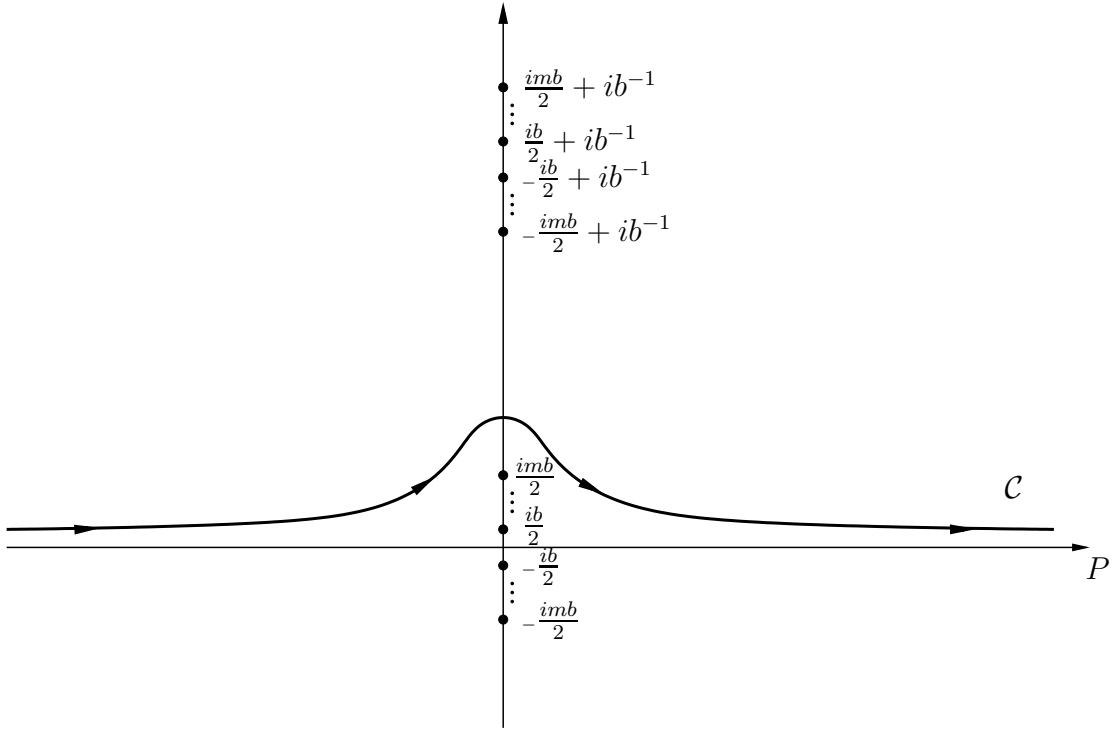}
	\caption{Contour of integration $\mathcal{C}$ in \eqref{4point-contour-integration}. This picture 
         is drawn with the assumption that $b^2\ll1$. Otherwise the contour has to be deformed.}
	\label{contour1}
\end{figure}

It is interesting that not only the conformal block has a simple expression for this special choice of the external conformal dimensions. The product of structure constants which enters in the definition of the four-point correlation function \eqref{OPE-4point} simplifies drastically and  is equal to\footnote{In order to simplify this product and obtain \eqref{CC} it is convenient to use the double argument formula for the $\Upsilon$-function \eqref{Upsilon-double-angle}.}
\begin{multline}\label{CC}
    C\left(-\frac{(2m-1)b}{4},\frac{Q}{2}-\frac{b}{4},\frac{Q}{2}+iP\right)
    C\left(\frac{Q}{2}-iP,\frac{Q}{2}-\frac{b}{4},\frac{Q}{2}-\frac{b}{4}\right)=16^{-2P^2}
    \Bigl[\pi\mu\gamma(b^2)b^{2-2b^2}\Bigr]^{\frac{m}{2}-\frac{1}{2b^2}}\times\\\times
    b^{2m+m(m+1)b^2}
    \frac{4\Upsilon(b)^2\Upsilon\left(\frac{(1-2m)b}{2}\right)
    \Upsilon\bigl(\frac{b}{2}\bigr)}{\Upsilon^2\bigl(\frac{b^{-1}}{2}\bigr)}
    \prod_{k=1}^m\gamma\left(ibP-\frac{kb^2}{2}\right)\gamma\left(-ibP-\frac{kb^2}{2}\right).
\end{multline}
As a consequence the integral over the intermediate momentum $P$ of the product of two structure constants \eqref{CC} with the modulus squared of the conformal block \eqref{Elliptic-block} which in fact gives the four-point correlation function can be performed analytically. We note that when the parameters $s_k$ of the fields take values \eqref{ak-sk} the dependence of the product of structure constants on the momentum $P$ can be expressed as a product of $2N$ $\gamma$-functions where $N$ is given by \eqref{N} depending linearly on this parameter. This fact supports our conjecture that in this case the solution is given by an integral of dimension $N$. The integration over the momentum $P$ is rather non-trivial mainly because the contour of integration is deformed as shown on figure \ref{contour1}. This deformation of the contour is prescribed by the condition that the four-point correlation function is single-valued. Surprisingly, the result of integration over the momentum $P$ is given by a multiple integral over the torus $T$ with periods $\pi$ and $\pi\tau$
\begin{multline}\label{4point-contour-integration}
   \int_{\mathcal{C}}\frac{|q|^{2P^2}\mathfrak{F}_m(P|\tau)\mathfrak{F}_m(-P|\tau^{*})}
   {\prod_{k=1}^m\sin\left(\pi(ibP+\frac{kb^2}{2})\right)\sin\left(\pi(ibP-\frac{kb^2}{2})\right)}\,dP=\\=
   \Lambda_m
   \Bigl(\text{Im}(\tau)\Bigr)^{-1/2}\int\limits_{T}\hspace*{-5pt}...\hspace*{-5pt}\int\limits_{T}
   \prod_{k=1}^m\mathcal{E}(u_k,\bar{u}_k)^{mb^2}
   \prod_{i<j}\mathcal{E}(u_i-u_j,\bar{u}_i-\bar{u}_j)^{-b^2}\,
   d^2u_1\dots d^2u_m,
\end{multline}
where
\addtocounter{equation}{-1}
\begin{subequations}
\begin{equation}
   \mathfrak{F}_m(P|\tau)\overset{\text{def}}{=}
   \int\limits_{0}^{\pi}\hspace*{-5pt}...\hspace*{-5pt}\int\limits_{0}^{\pi}e^{2bP(u_1+\dots+u_m)}
  \prod_{k=1}^{m}E(u_k)^{mb^2}\prod_{i<j}|E(u_i-u_j)|^{-b^2}\,du_1\dots du_m,
\end{equation}
\begin{equation}\label{Laplace-solution-torus}
     \mathcal{E}(u,\bar{u})=E(u)\bar{E}(\bar{u})e^{-\frac{2(\text{Im}u)^2}{\pi\text{Im}\tau}}
\end{equation}
and
\begin{equation}
   \Lambda_m=\frac{(-1)^m\,m!}{2^{m(m+1)b^2}}\frac
   {2^{2m-\frac{1}{2}}\sin^m\bigl(\frac{\pi b^2}{2}\bigr)}
   {\prod_{k=1}^m\sin\bigl(\frac{\pi kb^2}{2}\bigr)\,\sin\bigl(\frac{\pi (2m+1-k)b^2}{2}\bigr)}.    
\end{equation}
\end{subequations}
A phenomenon of this type is well known in the minimal models of CFT where the sum of squared modulus of conformal blocks which given by the contour Coulomb integrals taken with appropriate coefficients can be always expressed in terms of integrals over the plane and this property trivially ensure single-valuedness of this correlation function \cite{Dotsenko:1985hi}. Here we meet exactly the same situation but with the integral over intermediate the momentum $P$. This property was verified numerically with high accuracy (results of numerical calculations can be found on-line at \cite{url-Enrico}). We intend to give analytical proof in separate publication. The function $\mathcal{E}(u,\bar{u})$ given by \eqref{Laplace-solution-torus} is nothing but the exponent of the Green function of the Laplace-Beltrami operator on a torus
\begin{equation}
     \partial\bar{\partial}\log\left(\mathcal{E}(u,\bar{u})\right)=
     -\pi\delta^{(2)}(u)-\frac{1}{\pi\text{Im}\tau}.
\end{equation}
Finally we obtain the expression for the four-point correlation function 
\begin{multline}\label{4point-integral-representation}
   \langle V_{-\frac{(2m-1)b}{4}}(x,\bar{x})V_{\eta}(0)V_{\eta}(1)V_{\eta}(\infty)\rangle=
   \mathbb{C}_m
   \frac{\left|\frac{\theta_2(q)\theta_4(q)}{\theta_3(q)}\right|^{4m+2m(m+1)b^2}}
   {\left|x(x-1)\right|^{\frac{1}{2}+\frac{b^2}{4}}\left|\theta_3(q)\right|^2
   \left(\text{Im}(\tau)\right)^{1/2}}
   \times\\\times
   \int\limits_{T}\hspace*{-5pt}...\hspace*{-5pt}\int\limits_{T}
   \prod_{k=1}^m\mathcal{E}(u_k,\bar{u}_k)^{mb^2}
   \prod_{i<j}\mathcal{E}(u_i-u_j,\bar{u}_i-\bar{u}_j)^{-b^2}\,
   d^2u_1\dots d^2u_m,
\end{multline}
where 
\begin{equation*}
   \mathbb{C}_m=\frac{b^{m(m+1)b^2-2m}2^{6m+\frac{3}{2}}}{\pi^mm!}
   \frac{\Upsilon^2(b)\Upsilon^2\left(\frac{b}{2}\right)}
   {\Upsilon^2\left(\frac{b^{-1}}{2}\right)}
   \Bigl[\pi\mu\gamma(b^2)b^{2-2b^2}\Bigr]^{\frac{m}{2}-\frac{1}{2b^2}}
   \prod_{k=1}^m\frac{\gamma\left(-\frac{kb^2}{2}+\frac{1}{2}\right)}{\gamma\left(-\frac{kb^2}{2}\right)}.
\end{equation*}
The right hand side of \eqref{4point-integral-representation} up to a trivial factor looks like a Coulomb gas representation of the one-point correlation function of the operator $V_{-mb'}$ in LFT with parameter $b'=\frac{b}{\sqrt{2}}$ on a torus\footnote{On a torus the one-point correlation function of the operator $V_{-mb'}$ can be screened by $m$ screening fields, because curvature term does not contribute to the total charge in this case.}. We propose that this equality holds for general fields (i.e. for arbitrary not necessary integer values of $m$). Let us define function $\mathcal{T}(\alpha,b|q)$ related to the  one-point correlation function 
\begin{equation}
    \langle V_{\alpha}\rangle_{\tau}=
    \text{Tr}\left(q^{2L_0-\frac{c_L}{12}}\bar{q}^{2\bar{L}_0-\frac{c_L}{12}}V_{\alpha}(0)\right) 
\end{equation}
in Liouville field theory with cosmological constant $\mu$ and coupling constant $b$ on a torus with modular parameter $\tau$ as
\begin{equation}
      \mathcal{T}(\alpha,b|q)\overset{\text{def}}{=}
      \left[\pi\mu\gamma(b^2)b^{2-2b^2}\right]^{\frac{\alpha}{b}}
      |\eta(\tau)|^{-4\Delta(\alpha)}
      \langle V_{\alpha}\rangle_{\tau}
\end{equation}
where $\eta(\tau)=(\frac{1}{2}\theta_2(q)\theta_3(q)\theta_4(q))^{\frac{1}{3}}$ is Dedekind eta-function.
We define also the function $\mathcal{S}(\alpha,b|q)$ which is related to the four-point correlation function in LFT on sphere as
\begin{equation}
    \mathcal{S}(\alpha,b|q)\overset{\text{def}}{=}
    \left[\pi\mu\gamma(b^2)b^{2-2b^2}\right]^{\frac{\alpha}{b}+\frac{1}{2b}-\frac{1}{4}}
    \left|x(x-1)\right|^{\frac{4}{3}\Delta(\alpha)}
    \langle V_{\alpha}(x,\bar{x})V_{\eta}(0)V_{\eta}(1)V_{\eta}(\infty)\rangle.
\end{equation}
The correspondence between the one-point toric and the four-point spheric correlation functions states that
\begin{equation}\label{Torus-Sphere-duality}
     \mathcal{S}(\alpha,b|q)=\aleph\left(\bigl(\alpha-\frac{b}{4}\bigr)\sqrt{2},\frac{b}{\sqrt{2}}\right)\,
     \mathcal{T}\left(\bigl(\alpha-\frac{b}{4}\bigr)\sqrt{2},\frac{b}{\sqrt{2}}\Bigl|q\right),
\end{equation}
where $\aleph(\alpha,b)$ is given by
\begin{equation*}
    \aleph(\alpha,b)=\frac{\Upsilon_b(\alpha)}{\Upsilon_b\left(\frac{1}{2b}\right)}
    \frac{\Upsilon_b\left(\frac{1}{b}\right)}{\Upsilon_b\left(\alpha+\frac{1}{2b}\right)}.
\end{equation*}
We propose to give a proof of this relation for arbitrary values of the parameter $\alpha$ in a different publication. We note here that the relation between the conformal dimension of the operator on a torus (we denote it as $\Delta_{\tau}$) and the conformal dimension of the operator on a sphere (we denote it as $\Delta_S$) in \eqref{Torus-Sphere-duality} has the form
\begin{equation}
      \Delta_S=\frac{1}{2}\Delta_{\tau}+\frac{1}{4}+\frac{3b^2}{16}.
\end{equation}

In conclusion we note that all results of this section can be considered in the theory with $c<1$ ($b^2<0$). We did not discuss this problem here. We stress only that our results can be useful for the calculation of four-point amplitudes in Liouville gravity (see for example \cite{Belavin:2006ex,Kostov:2005av}). Not going into details, let us consider the analytical continuation of Liouville four-point correlation function \eqref{4point-integral-representation} to the region $b^2<0$, where it can be considered as a correlation function in matter theory with central charge $26-c_L$. The corresponding correlation function of dressing fields (fields with dimensions $\tilde{\Delta}_k=1-\Delta_k$, where $\Delta_k$ is conformal dimension of matter field) in Liouville theory  will be
\begin{equation}
    \langle V_{\frac{(2m+3)b}{4}}(x,\bar{x})V_{\eta}(0)V_{\eta}(1)V_{\eta}(\infty)\rangle,   
\end{equation}
i.~e. exactly the correlation function which is complementary to \eqref{4point-integral-representation} in a sense that it is described by the second term in the operator product \eqref{OPE-1/2b}. This second term defines  the conformal block 
\begin{equation}\label{Conf-Block-2}
  \tilde{\mathcal{H}}_m^{(P)}(q)\overset{\text{def}}{=}
  \mathfrak{H}_P\Bigl(\genfrac{}{}{0pt}{}{\frac{Q}{2}-\frac{b}{4}\;\;\;\;\;\;\frac{Q}{2}-\frac{b}{4}}
  {\frac{(2m+3)b}{4}\;\;\frac{Q}{2}-\frac{b}{4}}\Bigl|q\Bigr).
\end{equation}
which corresponds to the solution of \eqref{Diff-double-periodic-3} with asymptotic $\Psi\sim u^{m+1}$. This solution however has more complicated expression in terms of elliptic functions. For example the conformal block $\tilde{\mathcal{H}}_1^{(P)}(q)$ has the form
\begin{equation}
     \tilde{\mathcal{H}}_1^{(P)}(q)=\tilde{N}_1^{-1}\int_{0}^{\pi}e^{2bPu}E(u)^{b^2}\wp'(u)du,  
\end{equation}
where the normalization constant $\tilde{N}_1$ is given by
\begin{equation*}
     \tilde{N}_1=\frac{2^{4-b^2}\pi bPe^{\pi bP}}{(b^2-1)(b^2-2)}
     \frac{\Gamma(b^2)}{\Gamma\left(\frac{b^2}{2}-ibP\right)\Gamma\left(\frac{b^2}{2}+ibP\right)}.
\end{equation*}
The same is true in the more general situation when all $s_k$ is eq \eqref{Diff-double-periodic} are integer ($s_k=m_k$). Both correlation functions in matter and Liouville theory in this case have integral representations complementary to each other. We plan to study the corresponding four-point amplitudes in Liouville gravity elsewhere. 
\section{Concluding remarks}\label{Conclusion}
The differential equation defined in section \ref{DiffEq} has solution given by the integral \eqref{Liouville_4point}. Similar integrals appear in different applications in CFT, in supersymmetric CFT \cite{Belavin:2007gz} in CFT's having higher spin symmetry like Toda field theory \cite{Fateev:2007ab,Fateev:2008bm}, in different perturbed models of CFT etc. The results of the section \ref{DiffEq} give us a correspondence between the four parametric family of differential equations and the integrals \eqref{Liouville_4point}. These differential equations permit effectively to calculate the integrals \eqref{Liouville_4point} for all values of $x$.

Here we considered the family of the operators $V_{-\frac{mb}{2}}$ or $V_{-\frac{n}{2b}}$. For general case of the operator $V_{-\frac{mb}{2}-\frac{n}{2b}}$ the differential equation is of order $(m+1)(n+1)$ and can be written in the form \eqref{DIFF-n-x} but with functions $W^{(k)}(x)$ which are not anymore linear functions of the conformal dimensions $\Delta(\alpha_k)$ of the fields $V_{\alpha_k}$. We note however, that the solution to this equation can be represented as a bilinear superposition of the solutions to differential equations of orders $(m+1)$ and $(n+1)$. In particular in the simplest case $m=n=1$ the solution to the fourth order differential equation can be represented as a bilinear combination of hypergeometric functions. The correlation function for the field $V_{-\frac{b}{2}-\frac{1}{2b}}$ $=V_{-\frac{Q}{2}}(x,\bar{x})$ with three arbitrary fields up to a constant can be represented by the integral\footnote{This integral has a singularity at $t\rightarrow u$ and should be understood as a bilinear combination of contour integrals (see \cite{Dotsenko:1984nm}).}
\begin{equation}\label{4point-22}
  \langle V_{-\frac{Q}{2}}(x,\bar{x})V_{\alpha _{1}}\left( 0\right)
  V_{\alpha _{2}}\left( 1\right) V_{\alpha _{3}}\left( \infty \right) \rangle
  \rightarrow \left\vert x\right\vert ^{2\alpha _{1}Q}\left\vert
  x-1\right\vert ^{2\alpha _{2}Q}I(p_{1},p_{2},p_{3},x) 
\end{equation}
where
\begin{equation}
I(p_{1},p_{2},p_{3},x)=\int\left\vert t\right\vert^{2p_{1}}
                       \left\vert t-1\right\vert ^{2p_{2}}\left\vert t-x\right\vert^{2p_{3}}
                       \left\vert u\right\vert ^{2p_{1}^{\prime }}
                       \left\vert u-1\right\vert ^{2p_{2}^{\prime }}
                       \left\vert u-x\right\vert ^{2p_{3}^{\prime}}
                       \left\vert t-u\right\vert ^{-4}\,d^{2}t\,d^{2}u
\end{equation}
here $p_{1}=b(\alpha -2\alpha _{1}-Q/2)$, $p_{2}=b(\alpha -2\alpha_{2}-Q/2)$, 
$p_{3}=b(3Q/2-\alpha )$; with $\alpha =\alpha _{1}+\alpha_{2}=\alpha _{3}$ and $p_{i}^{\prime }/p_{i}=b^{-2}$.

It is convenient to use the notations $p_{12}=p_{1}+p_{2}$, $p_{123}=p_{1}+p_{2}+p_{3}$,
then a useful integral relation for $I(p_{1},p_{2},p_{3})$ takes the form:
\begin{equation}
I(p_{1},p_{2},p_{3},x)=C\left\vert x\right\vert ^{2+2p_{13}}\left\vert
x-1\right\vert ^{2+2p_{23}}H(p_{1},p_{2},p_{3},x) 
\end{equation}
where the constant 
\begin{equation*}
   C=-(1+b^{2})^{-2}\frac{\gamma (-p_{123})}{\gamma
 (-p_{1})\gamma (-p_{2})\gamma (-p_{3})}
\end{equation*}
and
\begin{equation}
    H(p_{1},p_{2},p_{3},x)=\hspace*{-2pt}\int\left\vert t\right\vert^{-2p_{1}-2}
    \left\vert t-1\right\vert ^{-2p_{2}-2}\left\vert t-x\right\vert^{-2p_{3}-2}
    \left\vert u\right\vert ^{2p_{1}^{\prime }-2}
    \left\vert u-1\right\vert^{2p_{2}^{\prime }-2}\left\vert u-x\right\vert ^{2p_{3}^{\prime }-2}
    \left\vert t-u\right\vert ^{2}\,d^{2}t\,d^{2}u.
\end{equation}
The last integral can be easily rewritten in terms of contour integrals which can be expressed in terms of hypergeometric functions. Four conformal blocks in this correlation function can be classified by the parameter $\alpha _{i}^{\prime }$ of the intermediate primary field, which takes
values: $\alpha _{1}^{\prime }=\alpha _{1}-Q/2$, $\alpha _{2}^{\prime }=\alpha_{1}+Q/2$, 
$\alpha _{3}^{\prime }=\alpha _{1}-Q/2+b$, $\alpha _{4}^{\prime}=\alpha _{4}-Q/2+1/b$. 
We normalize our conformal blocks $G_{\alpha_{i}^{\prime }}$ by the condition: 
\begin{equation*}
G_{\alpha _{i}^{\prime }}=x^{\Delta(\alpha _{i}^{\prime })-\Delta (\alpha _{1})-\Delta (-Q/2)}(1+c_{1}x+...).
\end{equation*}
It is convenient to introduce the functions:
\begin{equation}
  \begin{aligned}
    &\mathcal{F}_{1}\mathcal{(}p_{1},p_{2},p_{3},x)=F(1+p_{3},2+p_{123},2+p_{13},x),\\
    &\mathcal{F}_{2}\mathcal{(}p_{1},p_{2},p_{3},x)=x^{-1-p_{13}}F(1+p_{2},-p_{1},-p_{13},x),
  \end{aligned}
\end{equation}
where $F(a,b,c,z)$ is hypergeometric function, and to define 
\begin{equation}
   G_{\alpha_{i}^{\prime }}(x)=
   x^{1+p_{13}+\alpha _{1}Q}(1-x)^{1+p_{13}+\alpha _{1}Q}\mathcal{G}_{\alpha _{i}^{\prime }}(x)
\end{equation}
then:
\begin{equation}
   \mathcal{G}_{\alpha _{1}^{\prime }}(x)=\mathcal{F}_{2}\mathcal{(}p_{1},p_{2},p_{3},x)
   \mathcal{F}_{1}\mathcal{(-}p_{1}^{\prime}-1,-p_{2}^{\prime },-p_{3}^{\prime },x)
   -\frac{(1-p_{123}^{\prime })p_{1}}{(1-p_{13}^{\prime })p_{13}}\mathcal{F}_{2}
    \mathcal{(}p_{1}-1,p_{2},p_{3},x)
   \mathcal{F}_{1}\mathcal{(-}p_{1}^{\prime },-p_{2}^{\prime },-p_{3}',x)
\end{equation}
The function $\mathcal{G}_{\alpha _{2}^{\prime }}(x)$ can be derived from 
$\mathcal{G}_{\alpha _{1}^{\prime }}(x)$ by the substitution $p_{i}\rightarrow -p_{i}^{\prime }$
\begin{multline}
  \mathcal{G}_{\alpha _{3}^{\prime }}(x)=\frac{\left( 1+p_{123}\right)
  (1-p_{13}^{\prime })}{\left( p_{2}+p_{2}^{\prime }\right) }\mathcal{F}_{1}
  \mathcal{(}p_{1},p_{2},p_{3},x)
  \mathcal{F}_{1}\mathcal{(-}p_{1}^{\prime}-1,-p_{2}^{\prime },-p_{3}^{\prime },x)-\\
  -\frac{\left( 1-p_{123}^{\prime }\right) (1+p_{13})}{\left(p_{2}+p_{2}^{\prime }\right) }
  \mathcal{F}_{1}\mathcal{(}p_{1}-1,p_{2},p_{3},x)
  \mathcal{F}_{1}\mathcal{(-}p_{1}^{\prime},-p_{2}^{\prime },-p_{3}^{\prime },x)
\end{multline}
and
\begin{multline}
  \frac{p_{3}(p_{123}+p_{123}^{\prime })}{p_{13}(1-p_{13})(1+p_{13}^{\prime })}
  \mathcal{G}_{\alpha _{4}^{\prime }}(z)=\mathcal{F}_{2}\mathcal{(}p_{1}-1,p_{2},p_{3},x)
  \mathcal{F}_{2}\mathcal{(-}p_{1}^{\prime},-p_{2}^{\prime },-p_{3}^{\prime },x)-\\- 
  \mathcal{F}_{2}\mathcal{(}p_{1},p_{2},p_{3},x)
  \mathcal{F}_{2}\mathcal{(-}p_{1}^{\prime }-1,-p_{2}^{\prime },-p_{3}^{\prime },x).
\end{multline}
We note that in case when all four fields are $V_{-Q/2}$ the correlation function \eqref{4point-22} was studied  in \cite{Zamolodchikov:1986db} where it was expressed in terms of the hypergeometric functions.

In this paper we did not consider the application of elliptic conformal blocks and correlation functions to physical problems and two-dimensional quantum gravity. We note only that for $b^2<0$ the correlation function \eqref{4point-integral-representation} has a logarithmic behavior at coinciding points and probably can be used in physically interesting logarithmic CFT's. Correlation function \eqref{5-point} appears also in studying $SU(2)$ WZNW models \cite{Zamolodchikov:1986bd} as well as in their non-compact versions associated with $\hat{sl}(2)$ and $H_3^{+}$ \cite{Teschner:2001gi}. In these cases the variable $z$ plays role of an isotopic variable. The results of this paper can be applied to studying of correlation functions in these models. We propose to consider the application of correlation functions derived in this paper to the two-dimensional quantum gravity in a future publication.
\section*{Acknowledgment}
This work was supported, in part, by RBRF-CNRS grant PICS-09-02-91064. The work of A. L. was supported  by DOE grant DE-FG02-96ER40949, by  RBRF grant 07-02-00799-a, by the Russian Ministry of Science and Technology under the Scientific Schools grant 3472.2008.2 and by the RAS program "Elementary particles and the fundamental nuclear physics". A.L. and E.O. thank the Laboratoire de Physique Th\'eorique et Astroparticules Universit\'e Montpellier~II for hospitality.
\Appendix
\section{Useful formulae}\label{formulae}
\paragraph{$\Upsilon(x)$ function.} This function is defined by integral representation:
\begin{equation}\label{Upsilon-Integral}
  \log\Upsilon(x)=\int_{0}^{\infty}\frac{dt}{t}
    \left[\left(\frac{Q}{2}-x\right)^2e^{-t}-\frac
    {\sinh^2\left(\frac{Q}{2}-x\right)\frac{t}{2}}
    {\sinh\frac{bt}{2}\sinh\frac{t}{2b}} 
    \right].
\end{equation}
Double argument formula for the $\Upsilon$-function:
\begin{equation}\label{Upsilon-double-angle}
   \Upsilon(2x)=\frac{2^{4x(x-\frac{Q}{2})+1}}{\Upsilon(\frac{b}{2})\Upsilon(\frac{b^{-1}}{2})}
   \Upsilon(x)\Upsilon_b(x+b/2)\Upsilon(x+b^{-1}/2)\Upsilon(x+Q/2).
\end{equation}
Shift formula for the $\Upsilon$-function (here we put a lower index $b$ to $\Upsilon$):
\begin{equation}\label{shift-Upsilon}
   2^{x(x-(\frac{b}{2}+\frac{1}{b}))+\frac{1}{2}}
   \Upsilon_b\left(x\right)\Upsilon_b\left(x+\frac{b}{2}\right)=
   \frac{\Upsilon_b\left(b\right)\Upsilon_b\left(\frac{b}{2}\right)}
   {\Upsilon_{\frac{b}{\sqrt{2}}}\left(\frac{b}{\sqrt{2}}\right)}
   \Upsilon_{\frac{b}{\sqrt{2}}}\left(x\sqrt{2}\right).
\end{equation}
\paragraph{Theta function.} The theta function $\Theta_1(u|\tau)$ defined by\footnote{During this paper we drop sometimes the dependence of the function $\Theta_1(u|\tau)$ on $\tau$.}
\begin{equation}
    \Theta_1(u|\tau)=\sum_{n=-\infty}^{\infty}(-1)^{n-\frac{1}{2}}q^{(n+\frac{1}{2})^2}e^{(2n+1)iu}  
\end{equation}
is solution to the differential equation
\begin{equation}
   \partial_u^2\Theta_1(u|\tau)-\frac{4i}{\pi}\partial_{\tau}\Theta_1(u|\tau)=0.
\end{equation}
It satisfies the following quasi-periodicity relations 
\begin{equation}
  \begin{aligned}  
   &\Theta_1(u+\pi)=-\Theta_1(u),\\
   &\Theta_1(u+\pi\tau)=-q^{-1}e^{-2iu}\Theta_1(u),  
  \end{aligned}  
\end{equation}
and transforms as follows under the action of the modular group
\begin{equation}
   \begin{aligned}  
   &\Theta_1(u|\tau+1)=i^{\frac{1}{2}}\Theta_1(u|\tau),\\
   &\Theta_1\left(\frac{u}{\tau}\Bigl|-\frac{1}{\tau}\right)=(i\tau)^{\frac{1}{2}}e^{\frac{iu^2}{\pi\tau}}
   \Theta_1(u|\tau).  
  \end{aligned}
\end{equation}
For some purposes it is useful to define the function $E(u)$
\begin{equation}
     E(u)=\frac{\Theta_1(u)}{\Theta_1'(0)}=\sin(u)+4\sin^3(u)q^2+12\sin^3(u)q^4+O(q^6).
\end{equation}
The following differential relation will be useful in the appendix \ref{Andre-proof}
\begin{equation}\label{Lame-iden-1}
  \left(\partial_u\partial_v-\frac{2i}{\pi}\partial_{\tau}\right)\frac{E(u+v)}{E(u)E(v)}=0.
\end{equation}
This relation can be also rewritten as
\begin{equation}\label{Lame-iden-3}
     \frac{E''(u+v)}{E(u+v)}+\frac{E''(u)}{E(u)}+\frac{E''(v)}{E(v)}-2\frac{E'(u+v)}{E(u+v)}
     \left(\frac{E'(u)}{E(u)}+\frac{E'(v)}{E(v)}\right)+2\frac{E'(u)}{E(u)}\frac{E'(v)}{E(v)}
     -\frac{\Theta_1'''(0)}{\Theta_1'(0)}=0.
\end{equation}
Other theta-functions can be expressed through $\Theta_1(u)$ as
\begin{equation}
   \Theta_2(u)=\Theta_1\left(u+\frac{\pi}{2}\right),\quad
   \Theta_3(u)=e^{iu}q^{\frac{1}{4}}\Theta_1\left(u+\frac{\pi}{2}+\frac{\pi\tau}{2}\right),\quad
   \Theta_4(u)=-ie^{iu}q^{\frac{1}{4}}\Theta_1\left(u+\frac{\pi\tau}{2}\right).
\end{equation}
We define also theta constants
\begin{equation}
    \theta_2(q)=\Theta_2(0),\qquad
    \theta_3(q)=\Theta_3(0),\qquad 
    \theta_4(q)=\Theta_4(0),
\end{equation}
and functions
\begin{equation}\label{E_k}
      E_1(u)=E(u),\qquad
      E_2(u)=\frac{\Theta_2(u)}{\theta_2(q)},\qquad
      E_3(u)=\frac{\Theta_3(u)}{\theta_3(q)},\qquad
      E_4(u)=\frac{\Theta_4(u)}{\theta_4(q)}.
\end{equation}
\paragraph{Weierstra\ss\, function and Lam\'e equation.}
The Weierstra\ss\, function is defined by the infinite sum \eqref{Wp} and can be expressed through the second logarithmic derivative of the theta-function $\Theta_1(u)$ as
\begin{equation}
    \wp(u)=\left(\frac{\Theta_1'(u)}{\Theta_1(u)}\right)^2-\frac{\Theta_1''(u)}{\Theta_1(u)}+
    \frac{1}{3}\frac{\Theta_1'''(0)}{\Theta_1'(0)}.
\end{equation}
It is related to the elliptic sine function as
\begin{equation}
    \wp(u)=\left(\frac{2K(x)}{\pi}\right)^2\left(
     \sn^{-2}\left(\frac{2K(x)}{\pi}u\biggl|\sqrt{x}\right)-\frac{1}{3}(x+1)\right).
\end{equation}
The Weierstra\ss ~function has the expansion at the origin
\begin{equation}
   \wp(u)=\frac{1}{u^2}+\frac{g_2}{20}u^2+\frac{g_3}{28}u^4+O(u^6), 
\end{equation}
where the numbers $g_2$ and $g_3$ also known as invariants are given by
\begin{equation}
    \begin{aligned}
      &g_2=\frac{4}{3}\left(\frac{2K(x)}{\pi}\right)^4(x^2-x+1),\\
      &g_3=\frac{4}{27}\left(\frac{2K(x)}{\pi}\right)^6(x+1)(x-2)(2x-1).
    \end{aligned}
\end{equation}
We give also the double argument formula for the Weierstra\ss\, function
\begin{equation}
    \sum_{k=1}^4\wp(u-\omega_k)=4\wp(2u),  
\end{equation}
where $\omega_k$ are half periods given by \eqref{half-periods}.
The Lam\'e equation
\begin{equation}
     \left(-\partial_u^2+2\wp(u)\right)\Psi(u)=-\wp(v)\Psi(u)
\end{equation}
has a solution
\begin{equation}
     \Psi(u)=\frac{E(u+v)}{E(u)E(v)}e^{-\frac{\Theta_1'(v)}{\Theta_1(v)}u},
\end{equation}
which trivially leads to the following identity 
\begin{equation}\label{Lame-iden-2}
   \left[-\partial^2_u+2\frac{\Theta_1'(v)}{\Theta_1(v)}\partial_u+2\wp(u)\right]
   \frac{E(u+v)}{E(u)E(v)}=\left(\frac{\Theta_1''(v)}{\Theta_1(v)}-
   \frac{1}{3}\frac{\Theta_1'''(0)}{\Theta_1'(0)}\right)\frac{E(u+v)}{E(u)E(v)}.
\end{equation}
\section{Covariant differential operators}\label{Examples}
Here we give the first few examples of covariant differential operators defined in section \ref{DiffEq}:
\begin{equation}
       \mathfrak{D}^{(n+1)}=\partial^{n+1}+\frac{n(n+1)(n+2)}{6}W^{(2)}(x)\partial^{n-1}+\dots
\end{equation}
The coefficient before the derivative of the order $(n-k)$ are some graded differential polynomials of order $(k+1)$ in the functions (currents) $W^{(2)}(x)\dots W^{(k+1)}(x)$\footnote{Currents $W^{(k)}(x)$ have weight $k$ and derivative $\partial_x$ has degree $1$.}. Some coefficients of these polynomials are fixed from the condition that currents transform in appropriate way under the action of the diffeomorphism group
\begin{equation}
  \begin{gathered}
   x\rightarrow\omega(x),\\
   W^{(2)}(x)\rightarrow\Bigl(\frac{d\omega}{dx}\Bigr)^{2}W^{(2)}(\omega)+
   \frac{1}{2}\{\omega,x\},\qquad
   W^{(k)}(x)\rightarrow\Bigl(\frac{d\omega}{dx}\Bigr)^{k}W^{(k)}(\omega)
   \quad\text{for}\quad k>2.
  \end{gathered}
\end{equation} 
Namely, up to order $5$ all coefficients are fixed from this condition (up to total normalization) 
\begin{equation}\label{DIFF2}
   \mathfrak{D}^{(2)}=\partial^2+W^{(2)}(x),
\end{equation}
\begin{equation}\label{DIFF3}
  \mathfrak{D}^{(3)}=\partial^3+4W^{(2)}(x)\partial+2\partial W^{(2)}(x)+W^{(3)}(x),
\end{equation}
\begin{multline}\label{DIFF4}
  \mathfrak{D}^{(4)}=\partial^4+10W^{(2)}(x)\partial^2+\bigl(10\partial 
  W^{(2)}(x)+6W^{(3)}(x)\bigr)\partial+\\+
  \bigl(9W^{(2)}(x)^2+3\partial^2W^{(2)}(x)+3\partial W^{(3)}(x)+W^{(4)}(x)\bigr),
\end{multline}
\begin{multline}\label{DIFF5}
  \mathfrak{D}^{(5)}=\partial^5+20W^{(2)}(x)\partial^3+\bigl(30\partial 
  W^{(2)}(x)+21W^{(3)}(x)\bigr)\partial^2+\\+
  \bigl(64W^{(2)}(x)^2+18\partial^2W^{(2)}(x)+21\partial W^{(3)}(x)+8W^{(4)}(x)\bigr)\partial+\\+
  \bigl(32\partial\bigl(W^{(2)}(x)\bigr)^2+4\partial^3W^{(2)}(x)+6\partial^2W^{(3)}(x)+
  4\partial W^{(4)}(x)+48W^{(2)}(z)W^{(3)}(x)+W^{(5)}(x)\bigr).
\end{multline}
First ambiguity appears at level $6$, because one cannot distinguish between the functions $W^{(6)}(x)$ and $(W^{(3)}(x))^2$ using only their transformation properties. Both of them transforms like a tensor, so in order to separate them some other symmetries should be taken into account. Our analysis in section \ref{DiffEq} gives the differential operator $\mathfrak{D}^{(6)}$ and all other operators. For example $\mathfrak{D}^{(6)}$ equals
\begin{multline}\label{DIFF6}
  \mathfrak{D}^{(6)}=
  \partial^6+35W^{(2)}(x)\partial^4+\Bigl(70\partial W^{(2)}(x)+56W^{(3)}(x)\Bigr)\partial^3+\\+
  \Bigl(259W^{(2)}(x)^2+63\partial^2W^{(2)}(x)+84\partial W^{(3)}(x)+36W^{(4)}(x)\Bigr)\partial^2+\\+
  \Bigl(259\partial\bigl(W^{(2)}(x)\bigr)^2+28\partial^3W^{(2)}(x)+48\partial^2W^{(3)}(x)+
  36\partial W^{(4)}(z)+440W^{(2)}(x)W^{(3)}(x)+10W^{(5)}(x)\Bigr)\partial+\\
  +\Bigl(5\partial^4W^{(2)}(x)+10\partial^3W^{(3)}(x)+10\partial^2W^{(4)}(x)+
   5\partial W^{(5)}(x)+155W^{(2)}(x)\partial^2W^{(2)}(x)+
  130\left(\partial W^{(2)}(x)\right)^2+\\
  +220\partial(W^{(2)}(x)W^{(3)}(x))+225(W^{(2)}(x))^3+100W^{(2)}(x)W^{(4)}(x)+
  100(W^{(3)}(x))^2+W^{(6)}(x)\Bigr).
\end{multline}
As we see from eqs \eqref{DIFF2}-\eqref{DIFF6} the form of these differential operator is canonical in a sense that their coefficients are integer numbers which can be easily obtained from the WZW equation \eqref{WZ1} with finite matrices $J_{+}$ and $J_{-}$ given by \eqref{Js-finite}.
\section{Integral representation for solutions to the generalized Lam\'e heat equation}\label{Andre-proof}
We consider differential equation
\begin{equation}\label{Lame-appendix}
    \left[-\partial_u^2+\sum_{k=1}^4s_k(s_k+1)\wp(u-\omega_k)\right]\Psi(u|q)=
    \frac{4i}{\pi b^2}\partial_{\tau}\Psi(u|q).
\end{equation}
As was argued in the section \ref{Int-potentials} this equation is integrable for values of the parameters $s_k=m_k+\frac{2n_k}{b^2}$. We consider for simplicity the case $s_1=s_2=s_3=0$ and $s_4=1$. We have 
\begin{equation}
    \left[-\partial_u^2+2\wp(u)\right]\Psi(u|q)=
    \frac{4i}{\pi b^2}\partial_{\tau}\Psi(u|q).
\end{equation}
Let us try to find a solution in a form
\begin{equation}\label{Lame-anzatz-1}
    \Psi(u|q)=\int\rho(v|q)\frac{E(u+v)}{E(u)E(v)}\Psi_0(u+b^2v|q)dv,
\end{equation}
where we choose the integration limits in such a way that surface terms can be neglected\footnote{It will be justified below, that it can be done.}. Applying the differential operator $(-\partial_u^2+2\wp(u)-\frac{4i}{\pi b^2}\partial_{\tau})$ to the integral  \eqref{Lame-anzatz-1} and using identities \eqref{Lame-iden-1} and \eqref{Lame-iden-2} we find that the function $\rho(v|q)$ should satisfy two equations
\begin{equation}
   \frac{1}{b^2}\frac{\partial_v\rho(v|q)}{\rho(v|q)}=\frac{\Theta_1'(v)}{\Theta_1(v)},\qquad\qquad
   \frac{\Theta_1''(v)}{\Theta_1(v)}-\frac{4i}{\pi b^2}\frac{\partial_{\tau}\rho(v|q)}{\rho(v|q)}
   =\frac{1}{3}\frac{\Theta_1''(0)}{\Theta_1'(0)}
\end{equation}
with solution
\begin{equation}
     \rho(v|q)=\left(\frac{\Theta_1(v)}{\Theta_1'(0)^{\frac{1}{3}}}\right)^{b^2}.
\end{equation}
So the contour of integration goes between any two zeroes of $\Theta_1(v)$.

Generalization to the case $s_1=s_2=s_3=0$ and $s_4=m$ is very straitforward. We have
\begin{equation}
    \left[-\partial_u^2+m(m+1)\wp(u)\right]\Psi(u|q)=
    \frac{4i}{\pi b^2}\partial_{\tau}\Psi(u|q).
\end{equation}
The solution has the form
\begin{equation}\label{Lame-anzatz-m}
    \Psi(u|q)=\int\hspace*{-5pt}...\hspace*{-5pt}\int   
    \prod_{k=1}^m\left(\frac{\Theta_1(v_k)}{\Theta_1'(0)^{\frac{1}{3}}}\right)^{mb^2}
   \frac{E(u+v_k)}{E(u)E(v_k)}
   \prod_{i<j}\left|\frac{\Theta_1(v_i-v_j)}{\Theta_1'(0)^{\frac{1}{3}}}\right|^{-b^2}
   \,\Psi_0(u+b^2v|q)\,
   dv_1...dv_m,
\end{equation}
where $v=v_1+\dots+v_m$. It can be verified again with the help of relations \eqref{Lame-iden-1} and \eqref{Lame-iden-2}. Acting with the differential operator $\left[-\partial_u^2+m(m+1)\wp(u)-\frac{4i}{\pi b^2}\partial_{\tau}\right]$ on \eqref{Lame-anzatz-m}, using the trivial identity
\begin{equation}
   \partial_u\left(\frac{E(u+v_k)}{E(u)E(v_k)}\right)\partial_u\Psi_0(u+b^2v|q)=\frac{1}{b^2}
   \partial_u\left(\frac{E(u+v_k)}{E(u)E(v_k)}\right)\partial_{v_k}\Psi_0(u+b^2v|q)
\end{equation}
and integrating by part (we assume that integration limits in \eqref{Lame-anzatz-m} allow to do that) we obtain an integrand which is proportional to
\begin{multline}\label{andre-1}
    -(m-1)\sum_{k=1}^m\left(\frac{E''(v_k)}{E(v_k)}-\frac{1}{3}\frac{\Theta_1'''(0)}{\Theta_1'(0)}\right)
    +2(m-1)\sum_{k=1}^m\frac{E'(v_k)}{E(v_k)}\left(\frac{E'(u+v_k)}{E(u+v_k)}-\frac{E'(u)}{E(u)}\right)-\\-
    2\sum_{i<j}\left(\frac{E'(u+v_i)}{E(u+v_i)}-\frac{E'(u)}{E(u)}\right)
    \left(\frac{E'(u+v_j)}{E(u+v_j)}-\frac{E'(u)}{E(u)}\right)
    -2\sum_{j\neq k}\frac{E'(u+v_k)}{E(u+v_k)}\frac{E'(v_k-v_j)}{E(v_k-v_j)}+\\+
    \sum_{i<j}\left(\frac{E''(v_i-v_j)}{E(v_i-v_j)}-\frac{1}{3}\frac{\Theta_1'''(0)}{\Theta_1'(0)}\right)+
    m(m-1)\left(\left(\frac{E'(u)}{E(u)}\right)^2-\frac{E''(u)}{E(u)}
   +\frac{1}{3}\frac{\Theta_1'''(0)}{\Theta_1'(0)}\right).
\end{multline}
Expression \eqref{andre-1} can rewritten as a sum of two terms
\begin{multline}\label{andre-Aterm}
     A=-(m-1)\sum_{k=1}^m
     \Bigl[\frac{E''(u+v_k)}{E(u+v_k)}+\frac{E''(u)}{E(u)}+\frac{E''(v_k)}{E(v_k)}
     -2\frac{E'(u+v_k)}{E(u+v_k)}\left(\frac{E'(u)}{E(u)}+\frac{E'(v_k)}{E(v_k)}\right)+\\+
     2\frac{E'(u)}{E(u)}\frac{E'(v_k)}{E(v_k)}
     -\frac{\Theta_1'''(0)}{\Theta_1'(0)}\Bigr],
\end{multline}
and
\begin{multline}\label{andre-Bterm}
     B=\sum_{i<j}
     \Bigl[\frac{E''(v_i-v_j)}{E(v_i-v_j)}+\frac{E''(u+v_i)}{E(u+v_i)}+\frac{E''(u+v_j)}{E(u+v_j)}
     -2\frac{E'(v_i-v_j)}{E(v_i-v_j)}\left(\frac{E'(u+v_i)}{E(u+v_i)}-
      \frac{E'(u+v_j)}{E(u+v_j)}\right)-\\-2\frac{E'(u+v_i)}{E(u+v_i)}\frac{E'(u+v_j)}{E(u+v_j)}
     -\frac{\Theta_1'''(0)}{\Theta_1'(0)}\Bigr].
\end{multline}
In both sums \eqref{andre-Aterm} and \eqref{andre-Bterm} each term is equal to zero identically due to  \eqref{Lame-iden-3}.

One can easily find an integral representation for the solution to \eqref{Lame-appendix} in the more general case $s_1=s_2=s_3=0$ and $s_4=m+\frac{2n}{b^2}$. We give it without a proof (which is however straitforward and based on using identities \eqref{Lame-iden-1}, \eqref{Lame-iden-3} and \eqref{Lame-iden-2})
\begin{multline}\label{Lame-sol-mn}
     \Psi(u|q)=\left(\Theta_1'(0)\right)^{\frac{2n}{3}(1-\frac{2}{b^2})}
     \dashint\hspace*{-5pt}...\hspace*{-1pt}\dashint   
     \prod_{k=1}^m\left(\frac{\Theta_1(v_k)}{\Theta_1'(0)^{\frac{1}{3}}}\right)^{mb^2+2n}
     \hspace*{-6.25pt}\frac{E(u+v_k)}{E(u)E(v_k)}
     \prod_{k=1}^n\left(\frac{\Theta_1(v'_k)}{\Theta_1'(0)^{\frac{1}{3}}}\right)^{2m+\frac{4n}{b^2}}
     \hspace*{-6.25pt}\left(\frac{E(u+v'_k)}{E(u)E(v'_k)}\right)^{\frac{2}{b^2}}\cdot\\\cdot
     \prod_{i<j}\left|\frac{\Theta_1(v_i-v_j)}{\Theta_1'(0)^{\frac{1}{3}}}\right|^{-b^2}
     \prod_{i<j}\left|\frac{\Theta_1(v'_i-v'_j)}{\Theta_1'(0)^{\frac{1}{3}}}\right|^{-\frac{4}{b^2}}
     \prod_{i,j}\left(\frac{\Theta_1(v_i-v'_j)}{\Theta_1'(0)^{\frac{1}{3}}}\right)^{-2}
     \Psi_0(u+b^2v+2v'|q)\,d^mv\,d^nv',
\end{multline}
where $v=v_1+\dots+v_m$, $v'=v'_1+\dots+v'_n$, $d^mv=dv_1\dots dv_m$ and $d^nv'=dv'_1\dots dv'_n$. The integrand in \eqref{Lame-sol-mn} has a singularity at $v_i\rightarrow v_j'$ and should be correctly defined as a principal value similar to what was done in Ref.\cite{Dotsenko:1984ad}. One can obtain a solution to \eqref{Lame-appendix} with arbitrary $s_k=m+\frac{2n}{b^2}$ (not necessary $s_4$) and other $s_j=0$ just by the substitution $E(u+v_k)\rightarrow E_k(u+v_k)$, $E(u+v'_k)\rightarrow E_k(u+v'_k)$ and $E(u)\rightarrow E_k(u)$ in \eqref{Lame-sol-mn}, where $E_k(u)$ are defined by \eqref{E_k} with identification
\begin{equation}
     s_4\rightarrow E_1(u),\qquad
     s_1\rightarrow E_2(u),\qquad
     s_2\rightarrow E_4(u),\qquad
     s_3\rightarrow E_3(u).
\end{equation}
\section{Normalization integrals}\label{Fateev-integrals}
Here we give some integrals which can be used as normalization factors for
conformal blocks. We introduce the notations:
\begin{equation}
  D_{n}(u)=\prod\limits_{i<j}^{n}\left\vert 2\sin (u_{i}-u_{j})\right\vert
  ,\quad d^{n}u=du_{1}...du_{n},\quad g=-b^{2}/2. 
\end{equation}
The first normalization integral that appears in the normalization of conformal
blocks has the form:
\begin{equation}
  I_{n}(a,p,g)=\int\limits_{0}^{\pi }\hspace*{-5pt}...\hspace*{-5pt}\int\limits_{0}^{\pi}
  \prod_{k=1}^ne^{2pu_{k}}\left( 2\sin( u_{k}) \right) ^{a}D_{n}(u)^{2g}d^{n}u 
\end{equation}
This integral can be calculated exactly and is equal to:
\begin{equation}
I_{n}(a,p,g)=\pi ^{n}\prod\limits_{j=1}^{n}\frac{\Gamma (1+jg)}{\Gamma
\left( 1+g\right) }\frac{e^{\pi p}\Gamma (1+a-g+jg)}{\Gamma (1+\frac{a}{2}
-g+jg+\mathrm{i}p)\Gamma (1+\frac{a}{2}-g+jg-\mathrm{i}p)}
\end{equation}
For the our purposes we need the values of this integral for $
a=nb^{2}$, $g=-b^{2}/2$ and $p=bP$
\begin{equation}
 I_{n}(nb^{2},bP,-b^{2}/2)=\prod\limits_{j=1}^{n}\frac{\Gamma (1-\frac{
 jb^{2}}{2})}{\Gamma \left( 1-\frac{b^{2}}{2}\right) }\frac{\pi e^{\pi
 p}\Gamma (1+(2n+1-j)\frac{b^{2}}{2})}{\Gamma (1+\frac{jb^{2}}{2}+\mathrm{i}
 bP)\Gamma (1+\frac{jb^{2}}{2}-\mathrm{i}bP)}.
\end{equation}
The modification of this integral that also appears in the applications is:
\begin{equation}\label{fateev-int}
 \mathbf{I}_{n}(nb^{2},bP,-b^{2})=\int\limits_{0}^{\pi
 }\hspace*{-5pt}...\hspace*{-5pt}\int\limits_{0}^{\pi }\prod_{k=1}^ne^{2bPu_{k}}\left( 2\sin( u_{k})
 \right) ^{nb^{2}}\cot \left( u_{i}\right) D_{n}(u)^{-b^2}\,d^{n}u.
\end{equation}
This integral is equal to $\mathbf{I}_{n}=I_{n}(nb^{2},bP,-b^{2}/2)h_{n}(P),
$ where 
\begin{equation}
h_{n}(P)=\prod\limits_{j=1}^{n}\frac{2\mathrm{i}((n+1-2j)\frac{b^{2}}{2}+
\mathrm{i}bP)}{(2n+1-j)\frac{b^{2}}{2}}.
\end{equation}
Another normalization integral appears, when we consider conformal blocks given by integrals from $-\pi\tau$ to $\pi\tau$ avoiding zero clockwise for pure imaginary $\tau$ and take the limit $\tau\rightarrow i\infty$ (we did not consider them in this paper, but they still are important):
\begin{equation}
 J_{n}(a,p,g)=\int\limits_{-\infty }^{\infty }\hspace*{-5pt}...\hspace*{-5pt}\int\limits_{-\infty
 }^{\infty }\prod_{k=1}^ne^{2\mathrm{i}pu_{k}}\left( 2\cosh (u_k\right)
 )^{a}D_{n}(\mathrm{i}u)^{2g}\,d^{n}u
\end{equation}
This integral can be reduced to the Selberg integral and is equal to: 
\begin{equation}
J_{n}(a,p,g)=\prod\limits_{j=1}^{n}\frac{\Gamma (1+jg)}{\Gamma \left(
1+g\right) }\frac{\Gamma (-\frac{a}{2}-g+jg+\mathrm{i}p)\Gamma (-\frac{a}{2}
-g+jg-\mathrm{i}p)}{2\Gamma (-a+g-jg)}.
\end{equation}
In particular
\begin{equation}
J_{n}(nb^{2},bP,-b^{2}/2)=\prod\limits_{j=1}^{n}\frac{\Gamma (1-j\frac{b^{2}
}{2})}{\Gamma \left( 1-\frac{b^{2}}{2}\right) }\frac{\Gamma (-\frac{jb^{2}}{2
}+\mathrm{i}bP)\Gamma (-\frac{jb^{2}}{2}-\mathrm{i}bP)}{2\Gamma (-(2n+1-j)
\frac{b^{2}}{2})}
\end{equation}
The integral analogous to \eqref{fateev-int} in this case is
\begin{equation}
  \mathbf{J}_{n}(nb^2,bP,-b^2/2)=\int\limits_{-\infty }^{\infty
  }\hspace*{-5pt}...\hspace*{-5pt}\int\limits_{-\infty }^{\infty }\prod_{k=1}^ne^{2\mathrm{i}bPu_{k}}\left(
2\cosh (u_k\right) )^{nb^{2}}\tanh \left( u_{k}\right) D_{n}(\mathrm{i}u)^{-b^{2}}\,d^{n}u
\end{equation}
which is equal to: $\mathbf{J}_{n}(a,p,g)=\mathrm{i}^{n}h_{n}\left( P\right)
J_{n}(nb^{2},bP,-b^{2}/2).$

For some conformal blocks considered in the appendix \ref{Conf-blocks} (eq \eqref{E6} for $k=2$) normalization factors are defined by the integral:
\begin{equation}
\mathcal{J}_{n}(P)=\int\limits_{0}^{\pi }\hspace*{-5pt}...\hspace*{-5pt}\int\limits_{0}^{\pi}
  e^{4iPu/b}\left( 2\sin \left( u_{i}\right) \right)^{(4n-2)/b^{2}}
 (2\cos \left( u_{i}\right) )^{2/b^{2}}D_{n}(u)^{-4/b^2}\,d^{n}u
\end{equation}
This integral can be calculated only for this special relations between the
parameters and is equal:
\begin{multline}
  \mathcal{J}_{n}(P)=\cosh \left( \pi \left( \frac{P}{b}+\frac{in}{
 b^{2}}\right) \right) \frac{\Gamma (1+\frac{2n}{b^{2}}+\frac{2iP}{b}
 )\Gamma (1+\frac{2n}{b^{2}}-\frac{2iP}{b})}{\Gamma (1+\frac{n}{b^{2}
 }+\frac{iP}{b})\Gamma (1+\frac{n}{b^{2}}-\frac{iP}{b})}
 \times\\\times 
\prod\limits_{j=1}^{n}\frac{\Gamma (1-\frac{2j}{b^{2}})}{\Gamma
\left( 1-\frac{2}{b^{2}}\right) }\frac{\pi e^{\pi(i/b^{2}+2P/b)}
\Gamma (1+\frac{(4n-2j)}{b^{2}})}{\Gamma (1+\frac{2j}{b^{2}}+
i\frac{2P}{b})\Gamma (1+\frac{2j}{b^{2}}-\frac{2iP}{b})}
\end{multline}
All other normalization integrals for conformal blocks considered in the appendix \ref{Conf-blocks} can be
expressed in terms of the integrals $I_{n}(4n/b^{2},2P/b,-2/b^{2})$  and 
$I_{n}((4n-2)/b^{2},2P/b,-2/b^{2})$.
\section{Integrals for conformal blocks}\label{Conf-blocks}
Here we list elliptic conformal blocks which have explicit rather simple integral representation. As was proposed in section \ref{Int-potentials} elliptic conformal blocks which have explicit integral representation can be labeled by four integer numbers $m_k$ and $n_k$ for $k=1,2,3,4$ (where $s_k=m_k+\frac{2n_k}{b^2}$)
\begin{equation}\label{elliptic.conf.block-def}
  H_P^{\pm}\biggl(\genfrac{}{}{0pt}{}{(m_2,n_2)\;\;(m_3,n_3)}{(m_1,n_1)\;\;(m_4,n_4)}\biggl|q\biggr)
  \overset{\text{def}}{=}\mathfrak{H}_P\biggl(\genfrac{}{}{0pt}{}
  {\hspace*{-1pt}\frac{Q}{2}-\frac{(2m_2+1)b}{4}-\frac{n_2}{b}
  \hspace*{35pt}\frac{Q}{2}-\frac{(2m_3+1)b}{4}-\frac{n_3}{b}}
  {\frac{Q}{2}-\frac{(2m_1+1)b}{4}-\frac{n_1}{b}\mp\frac{1}{2b}
  \;\;\;\frac{Q}{2}-\frac{(2m_4+1)b}{4}-\frac{n_4}{b}}\biggl|q\biggr),
\end{equation}
where the conformal block in the r.h.s. of \eqref{elliptic.conf.block-def} is defined as in \eqref{Elliptic-Block-definition} with the identification of the points $x$, $0$, $1$ and $\infty$ as follows
\begin{equation}
  (m_1,n_1)\mapsto x,\qquad
  (m_2,n_2)\mapsto 0,\qquad
  (m_3,n_3)\mapsto \infty,\qquad
  (m_4,n_4)\mapsto 1.
\end{equation}
Blocks $H_P^{-}(\dots|q)$ are more complicated and we will not consider them in this appendix (see some discussions at the end of the section \ref{Int-potentials}). We give expressions for the conformal blocks  up to normalization which can be always expressed in terms of $\Gamma$-functions using integrals given in the appendix \ref{Fateev-integrals}. To reduce the integrals considered in this appendix to normalization integrals we should take into account that in the limit $q\rightarrow0$
\begin{equation}
   E_1(u)\rightarrow\sin(u),\qquad
   E_2(u)\rightarrow\cos(u),\qquad
   E_3(u)\rightarrow1,\qquad
   E_4(u)\rightarrow1,
\end{equation}
where the functions $E_k(u)$ are given by \eqref{E_k}. Below we give expressions for the conformal blocks \eqref{elliptic.conf.block-def} mostly in cases when only several of numbers $m_k$'s or $n_k$'s are non-zero. We adopt a notation where we will write only the parameters $m_k$'s or $n_k$'s which are non-zero (see below).  

Firstly we consider the situation when only some numbers $m_k$'s are non-zero. The number of integrations coinsides in this case with the number of gaps \eqref{g} in the potential \eqref{double-periodic-potential}.
In the case $m_k=m$ and other $m_j=0$ the conformal block is given by $m$-dimensional integral
\begin{equation}\label{block-m-0-0-0}
     H_P^{+}(m_k=m)\sim\int\limits_0^{\pi}\hspace*{-5pt}...\hspace*{-5pt}\int\limits_0^{\pi}
     \prod_{a=1}^me^{2bPu_a}E_1(u_a)^{mb^2}
     \frac{E_k(u_a)}{E_1(u_a)}\prod_{a<a'}|E_1(u_{aa'})|^{-b^2}\,d^mu.      
\end{equation}
where $d^mu=du_1...du_m$. Now we consider the situation when two numbers $m_k$ in \eqref{elliptic.conf.block-def} are non-zero.  For the case $m_1=m$, $m_k=m$ and other $m_j=0$ we have
\begin{equation}\label{block1-m-m-0-0} 
         H_P^{+}(m_1=m_k=m)\sim\int\limits_0^{\pi}\hspace*{-5pt}...\hspace*{-5pt}\int\limits_0^{\pi}
         \prod_{a=1}^me^{4bPu_a}(E_1E_k(u_a))^{2mb^2}
         \prod_{a<a'}|E_1E_k(u_{aa'})|^{-2b^2}\,d^mu,
\end{equation}
where we denote for shortness $E_1E_k(u)\overset{\text{def}}{=}E_1(u)E_k(u)$. 
For $m_1=0$, $m_i=m$, $m_j=m$ and other  $m_k=0$ the conformal block is given by
\begin{equation} 
         H_P^{+}(m_i=m_j=m)\sim\int\limits_0^{\pi}\hspace*{-5pt}...\hspace*{-5pt}\int\limits_0^{\pi}
         \prod_{a=1}^me^{4bPu_a}(E_1E_k(u_a))^{2mb^2}
         \frac{E_iE_j(u_a)}{E_1E_k(u_a)}
         \prod_{a<a'}|E_1E_k(u_{aa'})|^{-2b^2}\,d^mu,
\end{equation}
Now we consider the less trivial case of non-coinciding integers $m_1=m$, $m_k=m-1$ and  $m_i=m_j=0$ with $i,j\neq1\neq k$. The integral in this case will be $m$-dimensional
\begin{equation}\label{block-m-m-1}
      H_P^{+}(m_1=m,m_k=m-1)\sim\int\limits_0^{\pi}\hspace*{-5pt}...\hspace*{-5pt}\int\limits_0^{\pi}
      \prod_{a=1}^me^{4bPu_a}(E_1E_k(u_a))^{2mb^2}
      \Bigl(\frac{E_iE_j(u_a)}{E_1E_k(u_a)}\Bigr)^{b^2}
      \prod_{a<a'}|E_1E_k(u_{aa'})|^{-2b^2}d^mu.
\end{equation}
For the case when all numbers $m_k=m$ conformal block can be derived from eq \eqref{block-m-0-0-0} with $k=1$ by substitution $b\rightarrow2b$.

Now we consider situation when some numbers $n_k$'s in \eqref{elliptic.conf.block-def} are non-zero. In this case the dimension of the integral is always equal to $N=\sum_{j}n_{j}$. In the case $n_k=n$ and other $n_j=0$ the conformal block is given by a $n$-dimensional integral
\begin{equation}\label{E6}
  H_P^{+}(n_k=n)\sim\int\limits_{0}^{\pi}\hspace*{-5pt}...\hspace*{-5pt}\int\limits_{0}^{\pi}
  \prod_{a=1}^{n}e^{4b^{-1}Pu_a}E_1(u_a)^{\frac{4n}{b^2}}
  \left(\frac{E_k(u_a)}{E_1(u_a)}\right)^{\frac{2}{b^2}}
  \prod_{a<a'}|E_1(u_{aa'})|^{-\frac{4}{b^2}}\,d^nu.
\end{equation}
In the case $n_1=n$, $n_k=n$  one has 
\begin{equation}\label{block-dual1-m-m-0-0}
  H_P^{+}(n_1=n_k=n)\sim\int\limits_{0}^{\pi}\hspace*{-5pt}...\hspace*{-5pt}\int\limits_{0}^{\pi}
  \prod_{a=1}^{2n}e^{4b^{-1}Pu_a}(E_1E_k(u_a))^{\frac{4n}{b^2}}
  \prod_{a<a'}|E_1E_k(u_{aa'})|^{-\frac{2}{b^2}}\,d^{2n}u.
\end{equation}
We note that the conformal block \eqref{block-dual1-m-m-0-0} is dual to the conformal block \eqref{block1-m-m-0-0} (with $m$ substituted by $2n$ and $b\rightarrow b^{-1}$ in \eqref{block1-m-m-0-0}). In the case $n_i=n$, $n_j=n$  for $i,j\neq1$ one has
\begin{equation}
  H_P^{+}(n_i=n_j=n)\sim\int\limits_{0}^{\pi}\hspace*{-5pt}...\hspace*{-5pt}\int\limits_{0}^{\pi}
  \prod_{a=1}^{2n}e^{4b^{-1}Pu_a}(E_1E_k(u_a))^{\frac{4n}{b^2}}
  \Bigl(\frac{E_iE_j(u_a)}{E_1E_k(u_a)}\Bigr)^{\frac{1}{b^2}}
  \prod_{a<a'}|E_1E_k(u_{aa'})|^{-\frac{2}{b^2}}\,d^{2n}u.
\end{equation}
This conformal block is dual to the conformal block \eqref{block-m-m-1} (with $m$ substituted by $2n$ and $b\rightarrow b^{-1}$). We give here examples of non-coinciding numbers $n_{j}$. For the case $n_{1}=n-1$, $n_{k}=n$  with other two numbers $n_{j}=0$ one has
\begin{equation}
  H_P^{+}(n_1=n-1,n_k=n)\sim\int\limits_{0}^{\pi}\hspace*{-5pt}...\hspace*{-5pt}\int\limits_{0}^{\pi}
  \prod_{a=1}^{2n-1}e^{4b^{-1}Pu_a}(E_1E_k(u_a))^{\frac{4n-2}{b^2}}
  \prod_{a<a'}|E_1E_k(u_{aa'})|^{-\frac{2}{b^2}}\,d^{2n-1}u.
\end{equation}
This block is dual to \eqref{block1-m-m-0-0} with $m\rightarrow2n-1$ and $b\rightarrow b^{-1}$.
It is interesting also to consider the case $n_1=n-1$ and $n_2=n_3=n_4=n$. The integral will be $(4n-1)$-dimensional
\begin{equation}
    H_P^{+}(n_1=n-1,n_2=n_3=n_4=n)
   \sim\int\limits_0^{\pi}\hspace*{-5pt}...\hspace*{-5pt}\int\limits_0^{\pi}
     \prod_{a=1}^{4n-1}e^{2b^{-1}Pu_a}E_1(u_a)^{\frac{4n-1}{b^2}}
     \prod_{a<a'}|E_1(u_{aa'})|^{-\frac{1}{b^2}}\,d^{4n-1}u,
\end{equation}
which is dual to the conformal block \eqref{block-m-0-0-0} with $m=4n-1$ and $k=1$. 
In the case $n_1=n_2=n_3=n_4=n$ a conformal block is given by the integral which is again dual to the conformal block \eqref{block-m-0-0-0} with $m=4n$ and $k=1$. We plan to discuss this intriguing duality elsewhere. 

We considered conformal blocks with only numbers  $m_k$'s either $n_k$'s not equal to nonzero. In the mixed case conformal blocks are given by more complicated integrals. For example
\begin{multline}\label{block-mn}
   H_P^{+}(m_k=m,n_k=n)\sim\dashint\limits_0^{\pi}\hspace*{-5pt}...\hspace*{-1pt}\dashint\limits_0^{\pi}   
   \prod_{a=1}^mE_1(u_a)^{mb^2+2n}\frac{E_k(u_a)}{E(u_a)}
   \prod_{a'=1}^nE_1(u'_{a'})^{2m+\frac{4n}{b^2}}
   \left(\frac{E_k(u'_{a'})}{E_1(u'_{a'})}\right)^{\frac{2}{b^2}}\cdot\\\cdot
   \prod_{a<c}|E_1(v_{ac})|^{-b^2}\prod_{a'<c'}|E_1(v'_{a'c'})|^{-\frac{4}{b^2}}
   \prod_{a,a'}(E_1(v_a-v'_{a'}))^{-2}\,d^mv\,d^nv'.
\end{multline}
Conformal block \eqref{block-mn} is originated from the integral \eqref{Lame-sol-mn} and hence the same regularization by taking the principal value should be applied.

In this appendix we considered only several examples of conformal blocks which have an integral representation. At present integral representations for conformal blocks corresponding to general values of integers $m_k$ and $n_k$ in \eqref{s_k} are unknown. At the end of this appendix we note that for all cases considered here the corresponding correlation functions can be derived in the same way as it was
done above (see eq \eqref{4point-integral-representation}). Namely we should change all elliptic theta functions 
$E_{i}(u)$ in the corresponding integrands by the functions 
\begin{equation}
    \mathcal{E}_{k}(u,\bar{u})=
     E_{k}(u)\bar{E}_{k}(\bar{u})e^{-\frac{2(\text{Im}u)^{2}}{\pi\text{Im}\tau}}
\end{equation}
multiply by $(\text{Im}\tau )^{-\frac{1}{2}}$,  put $P=0$ and perform the integrals in the variables $u,\bar{u}$ over a torus with periods $\pi$ and $\pi\tau$. 
\bibliographystyle{MyStyle} 
\bibliography{MyBib}
\end{document}